# Chemical Kinetics for *Operando* Electron Microscopy of Catalysts: 3D Modeling of Gas and Temperature Distributions During Catalytic Reactions

Joshua L. Vincent[1], Jarod W. Vance[1], Jayse T. Langdon[1,2], Benjamin K. Miller[1,3], and Peter A. Crozier[1]*

[1]*School for Engineering of Matter, Transport, and Energy, Arizona State University, Tempe, Arizona 85281*

[2]*Present address: Department of Mechanical Engineering, University of Texas at Austin, Austin, Texas 78712*

[3]*Present address: Gatan, Inc., Pleasanton, CA, USA.*

*Corresponding author email: crozier@asu.edu

# Abstract

*In situ* environmental transmission electron microscopy (ETEM) is a powerful tool for observing structural modifications taking place in heterogeneous catalysts under reaction conditions. However, to strengthen the link between catalyst structure and functionality, an *operando* measurement must be performed in which reaction kinetics and catalyst structure are simultaneously determined. To determine chemical kinetics for gas-phase catalysis, it is necessary to develop a reliable chemical engineering model to describe catalysis as well as heat and mass transport processes within the ETEM cell. Here, we establish a finite element model to determine the gas and temperature profiles during catalysis in an open-cell *operando* ETEM experiment. The model is applied to a $SiO_2$-supported Ru catalyst performing CO oxidation. Good agreement is achieved between simulated compositions and those measured experimentally across a temperature range of 25 – 350 °C. In general, for lower conversions, the simulations show that the temperature and gas are relatively homogeneous within the hot zone of the TEM holder where the catalyst is located. The uniformity of gas and temperature indicates that the ETEM reactor system behavior approximates that of a continuously stirred tank reactor. The large degree of gas-phase uniformity also allows one to estimate the catalytic conversion of reactants in the cell to within 10% using electron energy-loss spectroscopy. Moreover, the findings indicate that for reactant conversions below 30%, one can reliably evaluate the steady-state reaction rate of catalyst nanoparticles that are imaged on the TEM grid.

**List of symbols**:

| Symbol | Description |
|---|---|
| Re | Reynolds number |
| Ma | Mach number |
| Ra | Rayleigh number |
| $\nabla$ | Del operator |
| $\rho$ | Density |
| $\boldsymbol{u}$ | Velocity vector |
| $t$ | Time |
| $p$ | Pressure |
| $\mu_i$ | Viscosity of component $i$ |
| $n$ | Number of moles |
| $\bar{M}$ | Mean molar mass |
| $V$ | Volume |
| $R$ | Gas constant |
| $T$ | Temperature |
| $\epsilon$ | Emissivity |
| $D_{ik}$ | Maxwell-Stefan diffusion coefficient between components $i$ and $k$ |
| $D_i^m$ | Mixture-averaged diffusion coefficient for component $i$ |
| $\omega_i$ | Mass fraction of component $i$ |
| $x_i$ | Mole fraction of component $i$ |
| $\boldsymbol{j}_i$ | Mass diffusion flux vector of component $i$ |
| $R_i$ | Rate of consumption or disappearance of component $i$ due to a reaction |
| $\boldsymbol{N}_i$ | Overall mass flux vector for component $i$ |
| $M_n$ | Mass-averaged molar mass |
| $M_i$ | Molar masses of each component $i$ |
| $D_i^K$ | Knudsen diffusion coefficient for component $i$ |
| $\lambda$ | Molecular mean free path |
| $D_i^{mK}$ | Knudsen-corrected, mixture-averaged diffusion coefficient for component $i$ |
| $\varepsilon$ | Porosity |
| $\tau$ | Tortuosity |
| $P$ | Permeability |
| $D_{e,ik}$ | Effective Maxwell-Stefan interdiffusion coefficient for components $i$ and $k$ |
| $D_{e_i}^{mK}$ | Effective Knudsen-corrected, mixture-averaged diffusion coefficient of component $i$ |
| $k_i$ | Adjusted thermal conductivity for component $i$ |
| $k_{0,i}$ | Unadjusted thermal conductivity for component $i$ |
| $w_d$ | Wall separation distance |
| $\alpha$ | Thermal conductivity adjustment function fitting parameter |
| $C_{p,i}$ | Heat capacity at constant pressure for component $i$ |
| **q** | Conductive heat flux vector |
| $Q$ | Volumetric heat source |
| $\Delta H_{rxn}$ | Enthalpy of reaction |
| $r_i$ | Rate of reaction of component $i$ |
| $v_i$ | Stoichiometric coefficient of component $i$ |
| $A$ | Pre-exponential or frequency factor |
| $E_a$ | Activation energy |
| $[i]$ | Concentration of component $i$ |
| $A_0$ | Intrinsic pre-exponential factor |
| $d$ | Distance into *operando* pellet from outer surface |
| $\gamma$ | Factor controlling pellet loading uniformity |
| $X_{CO}$ | Conversion of CO |
| $\dot{n}_{CO,in}$ | Molar flow rate of CO into the ETEM reactor |
| $\dot{n}_{CO,out}$ | Molar flow rate of CO out of the ETEM reactor |
| $\dot{n}_{total,out}$ | Total molar flow rate of the ETEM reactor |
| $y_{CO_2}$ | Mole fraction of $CO_2$ |
| $y_{CO}$ | Mole fraction of CO |
| $r_1$ | Rate of $CO_2$ formation integrated throughout entire *operando* pellet |
| $r_2$ | Rate of $CO_2$ formation estimated through EELS conversion measurement |

| $r_3$ | Rate of CO$_2$ formation integrated at surface along inner hole of pellet |

# 1. Introduction

Heterogeneous catalysis is an important approach in the generation of value-added products or in the elimination of environmental pollutants, representing a global market that exceeds $25 billion [1]. The rational design of more active and stable catalysts requires a fundamental understanding of the atomic-scale structures that regulate functionality for the reaction of interest. However, deriving structure-activity relationships for high-surface area catalysts remains challenging, since catalytically significant atomic structures are dynamic entities known to emerge *during catalysis.* The active structure(s) responsible for catalytic functionality remains poorly characterized on many high surface area catalysts. The purpose of all *in situ* and *operando* investigations is to develop a fundamental understanding of the structures that form while catalysis is taking place. To understand catalytic functionality, an aspirational goal is to establish a link between the structure of the catalyst and chemical kinetics of the catalytic reaction.

*In situ* environmental transmission electron microscopy (ETEM) is a powerful tool for studying high surface area catalysts under reaction conditions. Modern instruments offer spatial resolutions better than 0.1 nm, and the ability to visualize atomic-scale dynamics has delivered fundamental insight into processes underlying catalyst activation [2], active metal sintering [3–6], metal-support interactions [7–9], surface reconstructions [10,11], and phase transformations [12,13], to name a few. For reviews of seminal research in applying ETEM to catalysis, see, for example, [14–18]. *In situ* approaches are currently still not able to reproduce the high pressures present in many commercial industrial gas-phase reactors which often run at 10 – 100 atmospheres. The open cell ETEM reactor investigated here typically runs at pressures on the order of $10^{-3}$ atmospheres. Windowed cells can achieve substantially higher pressures (up to a few atmospheres) but may still be several orders of magnitude away from industrial conditions [19]. The ability to extrapolate observations made from *in situ* TEM (or any other *in situ* characterization tool) to higher pressures will depend in part on the pressure sensitivity of the reaction and the detailed mechanism of the reaction pathway on the catalysts.

Moreover, reaction conditions inside electron microscopes (either the open cell described here or the windowed cell) will always be different from lab-scale plug flow reactors typically used to characterize and measure catalyst performance. In a plug-flow reactor there is usually a significant variation in gas composition along the reactor bed as reactants get converted to products. There is a continuum of conditions within the reactor and the conversion that is measured is an ensemble property arising from the billions of nanoparticles all seeing different gas environments. The heat and mass transport characteristics in both ETEM reactor or TEM windowed cell reactor will also be very different from the typical lab scale or industrial scale reactor. Consequently, the catalysts in the TEM may be more or less active compared to *ex situ* conditions, active sites may be poisoned (which may not be obvious in a TEM image), and the catalyst may also show different selectivities to product formation. An *in situ* TEM experiment will almost always show atomic level changes in the catalyst structure upon exposure to reaction conditions, but it may not always be clear how such changes relate to catalytic functionality – unless the concentration of catalytic product species is measured simultaneously in the microscope.

To address this limitation and elucidate catalytic structure-activity relationships, it is necessary to develop *operando* methods which will ideally provide quantitative data on chemical kinetics of product formation, *i.e.*, the activity and selectivity of the catalyst, simultaneously with atomic-resolution structural information. Developing quantitative *operando* TEM methods is the primary motivation for the current work. It is useful to discuss the different information that is available from a simple *in situ* versus *operando* TEM experiment. An *operando* experiment is also an *in situ* experiment, so for clarity, we refer to an experiment that only exposes a catalyst to reactant/product gases without confirming that catalysis is taking place as an *in situ* experiment. Simple *in situ* TEM characterizes the atomic structure of the catalyst under reaction conditions, whereas *operando* TEM characterizes both the catalyst structure and the *kinetics of catalysis* under reaction conditions. In general, a catalyst will show surface structural changes in the presence of reactants/products, but many of the structural motifs may not be relevant to catalytic functionality; they are simple spectator species formed due to gas-solid interactions. For example, exposing a Ru nanoparticle to reactants during CO oxidation (the reaction explored here) results in formation of spectator surface oxide domains which *lower* the activity of the catalyst effectively covering active sites [20]. Differentiating spectator species from catalytically relevant motifs is challenging but, in favorable cases, it can be addressed by correlating changes in surface structures with changes in chemical kinetics. Therefore, the ability to correlate changes in kinetics with changes in catalyst structure is critical to determine the active structural motif in the electron microscope. To move from *in situ* to *operando* approaches, it is necessary to understand the chemical engineering aspects of the electron microscope reactor so that quantitative chemical kinetics can be determined.

For an open-cell environmental TEM, modified specimen preparation methods have been developed which increase the volume of catalyst in the microscope and thus enable quantitative determination of the gas phase products of a catalytic reaction while providing a stable platform for atomic-resolution imaging [21,22]. In this approach, catalyst particles are dispersed on a porous glass fiber pellet and an inert metal grid. Both the pellet and the grid are loaded into a standard furnace-style heating holder. The much larger mass of catalyst on the pellet produces detectable catalytic conversions, and catalyst particles on the metal grid are imaged. In this *operando* pellet approach, the composition of product gases in the ETEM can be monitored with residual gas analysis (RGA) and electron energy-loss spectroscopy (EELS) [23,24].

To determine chemical kinetic parameters (e.g., steady-state reaction rates and activation energies) from catalytic conversion data, it is necessary to establish an appropriate reactor model for the system. This requires understanding the spatial distribution of reactant and product gases in the cell as well as temperature profiles during catalysis. The *operando* pellet reactor geometry differs greatly from reactor architectures typically employed in chemical reaction engineering, making it necessary to develop heat and mass transport models for the ETEM cell. Mortensen et al. have conducted computational fluid dynamics (CFD) simulations to investigate the temperature distribution in a furnace-type holder as a function of gas pressure and composition in an ETEM [25]. Their model gives a good description of the temperature variation in the sample for single gases of different pressures and thermal conductivity, but it does not address the issues of catalysis or multi-component mass transport, which are essential to determining chemical kinetics.

In this work, we develop a model to explore mass transport, chemical conversion, and heat transfer relevant to running an *operando* catalysis experiment in an open cell ETEM. We develop a CFD model of the ETEM with *operando* pellet reactor and determine the gas composition and temperature distributions for experimental conditions of catalysis. As an exemplary case study, the model is applied to a $SiO_2$-supported Ru catalyst performing CO oxidation ($CO + \frac{1}{2}O_2 \rightarrow CO_2$). Simulations were conducted with the commercial program COMSOL Multiphysics®. The Computational Fluid Dynamics, Heat Transfer, and Chemical Reaction Engineering modules were used to simulate the gas and temperature profiles during catalysis.

## 2. Materials and Methods: Computational Fluid Dynamics Model

*2.1. Design of ETEM Cell*

For this work, a FEI Titan ETEM environmental cell and the *operando* pellet reactor were modelled with the COMSOL Multiphysics® finite-element simulation software. Dimensions for most components were derived from manufacturer blueprints for the FEI Tecnai ETEM, which is geometrically similar to the Titan instrument. Since some dimensions were proprietary and not known exactly, the sensitivity of the results to small changes in the geometry was investigated (see Supplemental Appendix 1), which found insignificant effects that should not impact the overall behavior or interpretation of the model. The *operando* pellet dimensions are based on those used in our lab for *operando* TEM experiments [22]. The pellet should be large enough to allow the typical catalyst loading in the TEM to be increased by a factor of 100 compared to a typical TEM grid in order to give catalytic conversions that are large enough to be detected by EELS. The radius of the hole in the center of the pellet is set to 0.5 mm. This is a compromise between maximizing the viewing area of the TEM grid versus maintaining uniform gas composition across the pellet and the grid. Making the hole much smaller than 0.5 mm may also lead to significant charging effects.

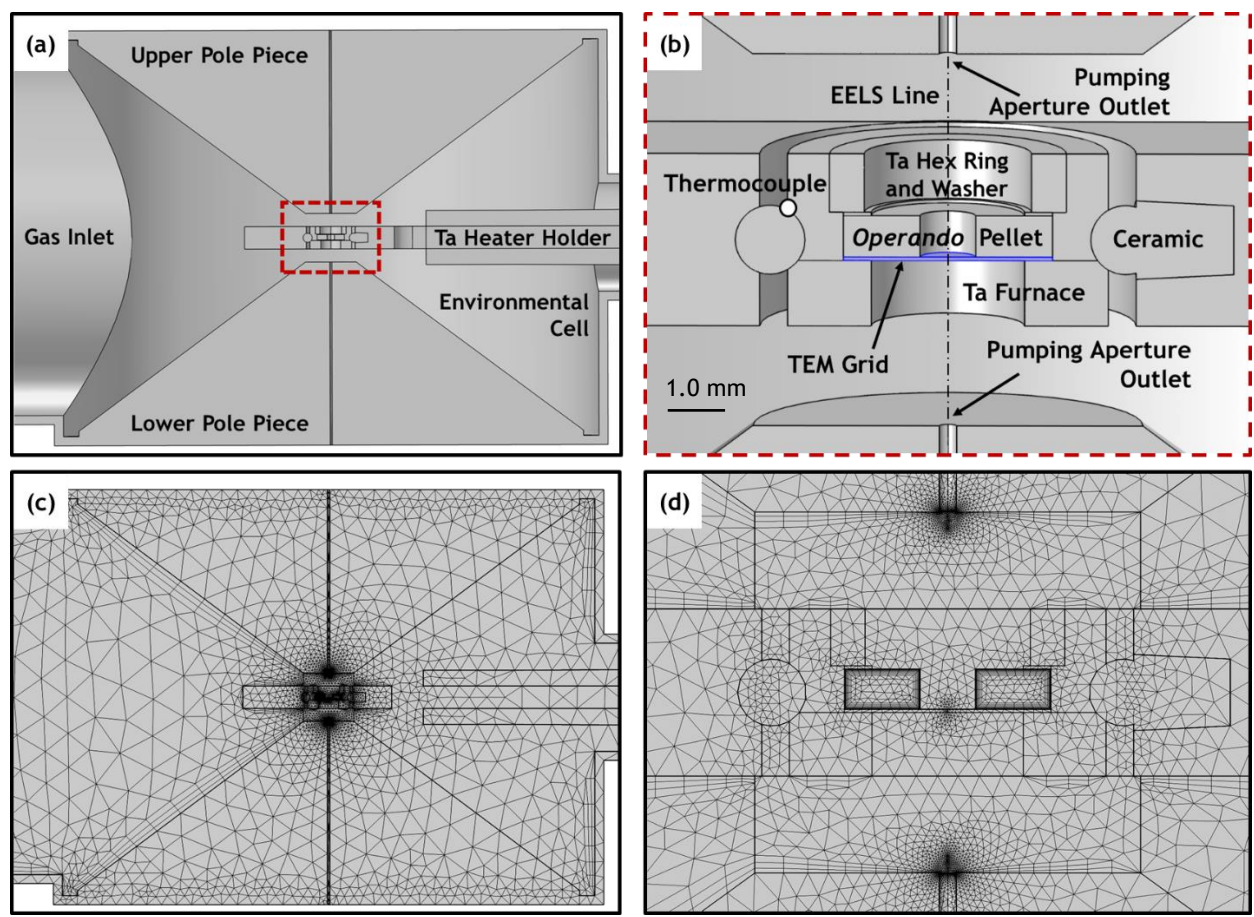

**Figure 1.** Environmental cell model geometry and finite element mesh **(a, c)** in full view, and **(b, d)** in an enhanced view focused on the *operando* pellet reactor. The TEM grid domain is colored blue for clarity.

The model geometry is shown in the top half of Figure 1, which displays (a) the environmental cell and (b) the *operando* pellet reactor. Note that the model takes advantage of mirror symmetry, and so only half of the chamber needs to be solved. A defined pressure and composition of reactant gas flows into the cell from the inlet on the left. The inlet extends approximately one meter out of the cell so the gas composition in the cell does not impact the composition at the inlet (see Supplemental Figure S1). Differential pumping apertures in the pole pieces serve as outlets. The *operando* pellet reactor and furnace holder reside between the pole pieces. A thermocouple on the outer edge of the Ta furnace is used to control the temperature of the reactor. The temperature and composition may be determined at any element in the model or averaged/integrated over chosen domains. The gas composition measured experimentally is replicated as an integral of the composition along the path labeled "EELS Line". The pressure outside the cell immediately beyond the differential pumping apertures is on the order of $2 \times 10^{-3}$ Torr [25]. For the typical *operando* TEM conditions discussed below, the pressure inside the cell is around 1,000 times higher. Therefore, nearly all of the experimental EELS signal arises from the gas localized to the high pressure zone within the volume of the cell, and it is sufficient to calculate the composition along the line marked in Figure 1b, rather than an extended path that traverses beyond the differential pumping aperture outlets. The composition at the grid is found through a surface integral on the plane labeled "TEM Grid," and the composition in the pellet is calculated as an integral over its volume. Comparing the gas and temperature distributions at these locations will yield insight into the prevalence and any potential effect of gradients that may exist during *operando* TEM experiments.

The gas and temperature distributions are computed by solving the relevant transport and chemical kinetic equations at many discrete elements throughout the geometry. Together, the set of all elements is known as a mesh. The mesh for this model is displayed in the bottom half of Figure 1, which shows (c) the full view of the cell and (d) and enhanced view of the *operando* pellet reactor. Small elements are used in the confined regions around the pellet reactor and differential pumping aperture outlets, while larger elements are used in the spacious environmental cell and gas inlet. The mesh is comprised of approximately 205,000 elements as this number was sufficient to achieve adequate convergence. An analysis of the mesh quality shows that refining the mesh with an increased number of elements does not greatly alter the converged solution but does unacceptably extend the computation time (see Supplemental Appendix 2).

The components labeled in Figure 1 are listed in Table 1, along with the materials with which they are modeled. As will be shown later, the gas and temperature distributions are largely uniform within the cell except for the small region localized to the furnace reactor and its immediate (< 3 mm) surroundings. Thus, including additional components found outside of this region (such as a cold trap, RGA sniffer tube, and objective aperture barrel), will unlikely impact the steady-state gas and temperature distributions, so these elements are excluded from the model for simplicity. Additionally, this line of reasoning is supported by results from previously published studies (see e.g., [25]). Materials properties have been imported from the COMSOL® library where appropriate (e.g., 304 steel). In other cases, properties are retrieved from references cited throughout the text. Materials properties relevant to heat and mass transport are given throughout the text and summarized in Table 2.

**Table 1**
Model components and properties.

| Component | Material |
|---|---|
| Gas inlet | Gas (varies) |
| Environmental cell | Gas (varies) |
| Pole pieces | 304 steel |
| Chamber surface | 304 steel |
| Heater holder body | Brass |
| Heater holder furnace, washer, and hex nut | Tantalum |
| Ceramic bridges | Zirconia |
| *Operando* pellet | Glass, porous |
| TEM grid | Tantalum, porous |
| Differential pumping aperture outlets | -- |
| Thermocouple | -- |
| EELS Line | -- |

**Table 2**

Material properties relevant to heat and mass transport.

| Material | Thermal Conductivity (W m$^{-1}$ K$^{-1}$) | Heat Capacity (J kg$^{-1}$ K$^{-1}$) | Emissivity | Viscosity (kg m s$^{-1}$) | Density (kg m$^{-3}$) | Diffusivity (cm$^2$ s$^{-1}$) |
|---|---|---|---|---|---|---|
| Steel | 44.5 | 475 | 0.3 | N/A, solid | 7,850 | N/A, solid |
| Tantalum | 57.5 | 140 | 0.05 | N/A, solid | 16,700 | N/A, solid |
| Zirconia | 2 | 500 | ~0 | N/A, solid | 5,700 | N/A, solid |
| Brass | 120 | 377 | 0.05 | N/A, solid | 8,500 | N/A, solid |
| Glass | 1.2 | 730 | 0.9 | N/A, solid | 2,210 | N/A, solid |
| CO | Equation S14 | Equation S11 | N/A, gas | Equation S4 | Ideal gas law | Equation S7 |
| O$_2$ | Equation S15 | Equation S12 | N/A, gas | Equation S5 | Ideal gas law | Equation S8 |
| CO$_2$ | Equation S16 | Equation S13 | N/A, gas | Equation S6 | Ideal gas law | Equation S9 |

## 2.2. *Treatment of Mass Transport*

For multi-component gas mixtures, mass transport can occur by bulk convection and inter-species diffusion. Bulk convection is modeled here with the Navier-Stokes equation. Under the conditions simulated, the soft limit on the Mach number of Ma = 0.3 was not exceeded [26]. The gas is considered laminar, compressible, Newtonian, and unaffected by gravitational forces. A dimensionless number analysis supports these simplifications (see Supplemental Appendix 3). With these considerations, the form of the equation solved in the model becomes [26]:

$$(1) \quad \rho \frac{\delta \boldsymbol{u}}{\delta t} + \rho (\boldsymbol{u} \cdot \nabla \boldsymbol{u}) = \nabla \cdot [-p\mathbf{I} + \mu(\nabla \boldsymbol{u} + (\nabla \boldsymbol{u})^\text{T}) - \frac{2}{3}\mu(\nabla \cdot \boldsymbol{u})\mathbf{I}]$$

This equation is solved simultaneously along with the equation of continuity [26]:

$$(2) \quad \frac{\delta \rho}{\delta t} + \nabla \cdot (\rho \boldsymbol{u}) = 0$$

Where $\rho$ is the fluid density, $\boldsymbol{u}$ is the velocity vector, $t$ is time, $\nabla$ is the del operator, $p$ is the pressure, $\boldsymbol{I}$ is the identity matrix, $\mu$ is the dynamic viscosity, and T is the transposition operation.

Only steady-state solutions were computed, so the time derivatives were set to zero. The density of the gas was modeled with the ideal gas law [27]. Polynomial expressions for the viscosity of CO, $O_2$, and $CO_2$ were determined from published data [28–30] and are provided in Supplemental Appendix 5. The viscosity of mixtures of these gases were computed by Wilke's method [31].

In the *operando* pellet, the permeability and porosity of the microfibrous pore network is factored in to attenuate the driving force for flow. The permeability, $P$, has been calculated from the pellet's published Herzberg speed [32,33], which produces a value of $P = 1.5 \times 10^{-12}$ m$^2$. The porosity, $\varepsilon$, has been approximated by taking a ratio of the density of the pellet against that of borosilicate glass, which yields a value of $\varepsilon = 0.7$. Given that the porosity was estimated, a sensitivity analysis was done to assess the effect of slight changes to the porosity. For physically plausible variations in the range of $0.1 - 0.2$, there were not significantly different outcomes in the behavior or interpretation of the model; more details are provided in Supplemental Appendix 7.

A mass flow boundary condition on the inlet surface defines the composition and flow rate of gas into the ETEM cell. At the pumping aperture outlets, a Vacuum Pump boundary condition defines the static cell pressure that arises for a given inlet composition and flow rate. The pressure-flow relationship is described from experimental data acquired on the ETEM. See Supplemental Appendix 4 for more details, including a description of the experimental vacuum system configuration used in the data acquisition. The solid surfaces exposed to gas were modeled with a no-slip boundary condition, which specifies the fluid velocity at the surface to be equal to zero.

Inter-species diffusion is described in the model with a Maxwell-Stefan approach [34,35]. Thermal diffusion is ignored since the molecules are similar in size and mass [36]. Electric migration is also ignored assuming that the electron beam is blanked for sufficient time. Expressions for the binary diffusivities of CO, $O_2$, and $CO_2$ are found in the low-pressure limit and provided in Supplemental Appendix 5 [37]. A simplification is implemented that averages the binary diffusivities against the local composition to produce a mixture-averaged diffusion coefficient for each component $i$, $D_i^m$. This mixture-averaged form is appropriate for gases of similar molecular weights (e.g., CO, $O_2$, and $CO_2$) and simplifies mass flux calculations by eliminating parameters that are difficult to compute [38,39].

With these simplifications, the overall mass transport equations may be written as:

(3) $\quad \nabla \cdot \boldsymbol{j_i} + \rho(\boldsymbol{u} \cdot \nabla)\omega_i = R_i$

(4) $\quad \boldsymbol{N_i} = \boldsymbol{j_i} + \rho\boldsymbol{u}\omega_i$

With supporting equations:

(5) $\quad \boldsymbol{j_i} = -\left(\rho D_i^m \nabla \omega_i + \rho \omega_i D_i^m \frac{\nabla M_n}{M_n}\right)$ (6) $\quad M_n = \left(\sum_i \frac{\omega_i}{M_i}\right)^{-1}$

Here, the symbols used in Equations (1) and (2) retain their same meaning. In Equations (3) and (4), $\boldsymbol{j_i}$ is the mass diffusion flux vector of component $i$, $\omega_i$ is the mass fraction of species $i$, $R_i$ is the rate of consumption or disappearance of species $i$ due to a reaction, and $\boldsymbol{N_i}$ is the overall mass flux vector for component $i$. In Equation (5), $D_i^m$ is the mixture-averaged diffusion coefficient for component $i$, and $M_n$ is the mass-averaged molar mass, which is computed in Equation (6) from the molar masses of each component $i$, $M_i$.

In the *operando* pellet, where diffusion is obstructed by the pellet's pore network, the diffusion coefficient is reduced with a Knudsen term as described in Equations (7) and (8) below:

(7) $\quad D_i^K = \frac{\lambda}{3}\sqrt{\frac{8RT}{\pi M_i}}$ (8) $\quad D_i^{mK} = \left(\frac{1}{D_i^m} + \frac{1}{D_i^K}\right)^{-1}$

In Equation (7), $D_i^K$ is the Knudsen diffusion coefficient for component $i$, $\lambda$ is the molecular mean free path, $R$ is the gas constant, and $T$ is the local temperature. The mean free path in the pellet was set to the pore diameter of 2.7 µm published by the manufacturer [32]. In Equation (8), the Knudsen-corrected, mixture-averaged diffusivity $D_i^{mK}$ is calculated by a parallel resistance treatment.

One final consideration is given to diffusion in the pellet, which further attenuates the diffusivities by factoring in the pellet's porosity, $\varepsilon$, and tortuosity, $\tau$. This operation is applied to the binary diffusivities before any calculations are done and is described by Equation (9) below:

(9) $\quad D_{e,ik} = \frac{\varepsilon}{\tau} D_{ik}$

Here, $D_{e,ik}$ is the effective binary diffusivity for components $i$ and $k$. As mentioned, the porosity has been evaluated as $\varepsilon = 0.7$. There are several methods for estimating tortuosity from porosity; in this work the Bruggeman method is chosen [40], yielding a value of $\tau = \varepsilon^{-1/2} = \sim 1.2$.

With the considerations for diffusion through the pore network implemented, the diffusive flux vector for species $i$ in the pellet becomes slightly modified to Equation (10) below:

(10) $\quad \mathbf{j}_i = -\left(\rho D_{e_i}^{mK} \nabla \omega_i + \rho \omega_i D_{e_i}^{mK} \frac{\nabla M_n}{M_n}\right)$

Here, $D_{e_i}^{mK}$ represents the effective mixture-averaged diffusion coefficient of component $i$ after correcting for Knudsen resistance to diffusion.

*2.3. Treatment of Heat Transport*

Heat transfer considerations include radiation, convection, and conduction. Radiation is treated with the Stefan-Boltzmann law [26]. Values of the emissivity for steel [41,42], brass [41], borosilicate glass [42], zirconia [41,43], and tantalum [41,44] have been located from the literature. Moderate errors in emissivity should have little impact on heat transfer for this model: in vacuum at 400 °C, which is the upper temperature investigated here, radiation only accounts for around 10% of the heat lost from the furnace, and at 250 °C it is less than 1%. Since radiation scales with temperature to the fourth power, radiation may become important above 500 °C. Heat transfer by convection is insignificant as well, given the low pressures/flow rates and small geometries present in the cell. A dimensionless number analysis of the Rayleigh number supports this conclusion (see Supplemental Appendix 3) [26]. Conduction is the most important mode of heat transport in this model. Conduction through solids is modeled with Fourier's law [26]. Thermo-physical properties for solid densities, heat capacities, and thermal conductivities have been gathered from tabulated sources [45–52] or imported from the COMSOL material library.

Conduction through gases is modeled by Fourier's law with a slight modification to the thermal conductivity of the gas. For macroscopic heat transfer, a gas's thermal conductivity is typically independent of pressure or of the geometry under consideration. Given the low pressures and confined spaces present in the ETEM reactor, the thermal conductivity may become dependent on these factors, as the mean free path between molecular collisions can become comparable to the wall distance that separates the surfaces transferring heat (usually, the molecular mean free path is many times smaller). In this case, gas-wall collisions may occur as frequently as gas-gas collisions, which effectively reduces the conductivity of the gas [53]. In the reactor the effect is relevant around the furnace and the pole pieces, where the wall separation distances are smallest.

A theoretical description of the pressure-distance dependence has not been found, so an empirical fit is employed, given by:

(11) $\quad k_i = k_{0,i} * \left(1 - \exp\left(\frac{-w_d}{\lambda * \alpha}\right)\right)$

Here, $k_i$ is the adjusted thermal conductivity for component $i$, $k_{0,i}$ is the unadjusted conductivity, $w_d$ is the distance to the nearest wall, $\lambda$ is the mean free path, and $\alpha$ is a fitting parameter. The mean free path was evaluated by the kinetic theory of gases [54]. The adjustment is implemented in two regions: (1) between the furnace and the pole pieces, and (2) between the furnace and the holder body. By inspection, one can see that the unadjusted conductivities are recovered in the limits of large $w_d$ (*i.e.,* unconfined spaces) or small $\lambda$ (*i.e.,* high pressure).

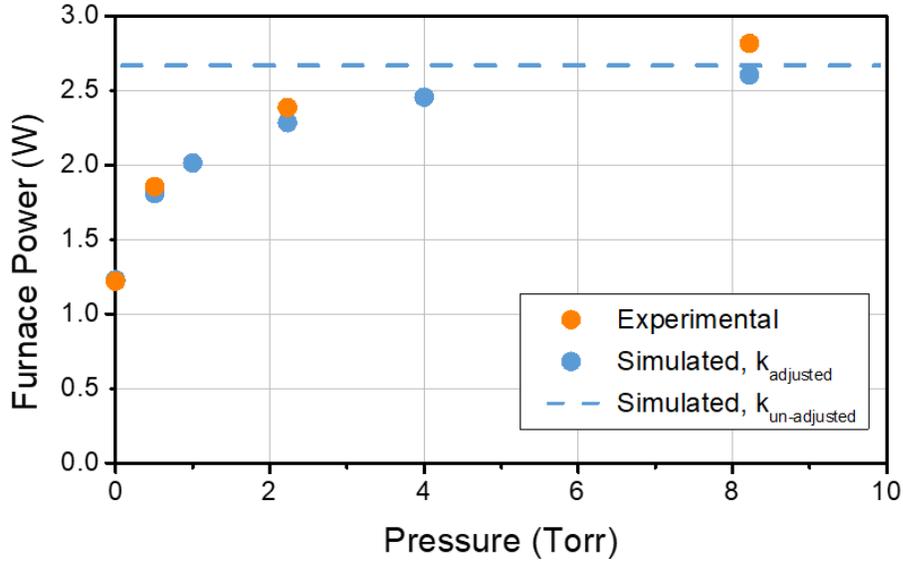

**Figure 2.** Furnace power versus pressure of CO. Power shown is that required to maintain the furnace at 400 °C. Simulated data is given with (blue circles) and without (dashed line) an adjustment to the thermal conductivity, showing that the adjustment provides a good fit to experimental data (orange circles).

The impact of the adjustment can be seen by examining the power required to maintain a constant furnace temperature in various pressures of gas. This is equivalent to the amount of heat transferred from the furnace to the surrounding gas. Figure 2 displays the power required to maintain a temperature of 400 °C in CO pressures ranging from $10^{-5}$ – 10 Torr. A temperature of 400 °C provides an upper bound on heat lost from the furnace. The required amount of power was measured experimentally and is plotted in orange circles. The flat, dashed blue line depicts the simulated heat lost when no conductivity adjustment is implemented. One can see that the unadjusted case shows no pressure dependence on power loss and thus disagrees badly with the experimental measurements shown in orange. When Equation (11) is implemented (blue circles), a good fit to the experimental behavior is obtained with a fitting parameter value of $\alpha = 15$. The vacuum simulation agrees very well with the experimental heat loss in vacuum (10$^{-5}$ Torr). Overall, these results are taken as evidence of a sufficiently descriptive heat transfer model.

With these considerations, the general energy balance and the equation for heat conduction, respectively, are given by Equations (12) and (13) below:

(12) $\quad \rho C_p \boldsymbol{u} \cdot \nabla T + \nabla \cdot \mathbf{q} = Q$

(13) $\quad \mathbf{q} = -k \nabla T$

Here, $C_p$ is the heat capacity, $\mathbf{q}$ is the conductive heat flux vector, and $Q$ is a volumetric heat source. The furnace is treated as a volumetric heat source. Enthalpic heat from the exothermic reaction is also treated as a volumetric heat source. The pole pieces are modeled as water cooled by setting their outer surfaces to 25 °C. The gas flowing into the cell is set to an initial temperature of 25 °C. Gases are considered to be ideal, with their heat capacity, enthalpy, and entropy taken as functions of temperature from ideal gas models [55]. Polynomial expressions for the heat capacities and unadjusted thermal conductivities of CO, $O_2$, and $CO_2$ are found as a function of temperature in the low-pressure limit [29,30,56] and are provided in Supplemental Appendix 5.

*2.4. Treatment of Chemical Reactions and Catalysis*

A Ru/SiO$_2$ catalyst performing CO oxidation (CO + ½O$_2$ → CO$_2$, and ΔH$_{rxn}$ = -283.0 kJ/mol) was chosen as a model system to determine the gas and temperature profiles during catalysis. The reaction is known to proceed with a negligible rate in regions of the cell where there is no catalyst, so the reaction is significant only in the *operando* pellet [21,57]. For simplicity, the reaction kinetics are modeled as elementary and irreversible [58], with rates for each species given by:

(14) $\quad r_i = v_i A \sqrt{T} e^{-\frac{E_a}{RT}} * [CO][O_2]^{\frac{1}{2}}$

Here $r_i$ is the rate of reaction of species $i$, $v_i$ is the stoichiometric coefficient of species $i$, $A$ is a pre-exponential factor, $T$ is the temperature, $E_a$ is the activation energy, $R$ is the gas constant, $[CO]$ is the concentration of CO, and [$O_2$] is the concentration of $O_2$. The general heat and mass transport behavior that emerges from the reactor model does not depend on the functional form of the kinetics chosen to describe the chemical reaction. In fact, as will be shown, the homogeneity in pressure and temperature around the catalyst implies that any kinetic equation will appear identical in the model if it produces the same reaction rate for each gaseous species at fixed temperature and partial pressure of reactants and products.

The spatial distribution of catalyst in the pellet was modeled with an egg-shell profile (see Supplemental Appendix 6; the profile used in the results presented below corresponds to that in Figure S5b). Experimentally, the total mass of catalyst is not known exactly (typically it is on the order of 20-200 µg, while the mass of the pellet is ~3000 µg), so in this work the value of A is adjusted until a match with experimental data is achieved for compositions measured at 340 °C (here, $A = 7*10^{12}$ s$^{-1}$). The adjustment is considered appropriate to account for uncertainties in catalyst mass, without making sacrifices to rigor in solving for or making conclusions about the resultant gas and temperature profiles. The $E_a$ of CO oxidation over Ru/SiO$_2$ has been measured, and the reported mean of 90 kJ mol$^{-1}$ was chosen [59,60].

## 2.5. *Discussion on Effect of Higher Operating Temperature (i.e., > 340 °C)*

Finally, it is worthwhile to briefly discuss the impact of higher operating temperature on the gas and temperature distributions in the cell and reactor. In general, to apply *operando* ETEM to various catalysts of interest, the furnace reactor temperature may need to exceed 340 °C (the highest temperature explored in the present study). The effect of higher specimen temperature can be understood by examining what happens to the dominant modes of mass and heat transport. As will be shown, given the broad uniformity of low-magnitude velocities throughout the cell, mass transport is dominated by diffusion and heat transfer by conduction. Increasing the reactor temperature will not change the dominant mode of mass transport, but it may make it more efficient, as the diffusivities which describe the intermixing of gas species become larger at higher temperatures (see, *e.g.,* Equations S7 – S9). In this sense the distribution of gas products in the cell and reactor could become more uniform for catalysts with higher light-off temperatures. For heat transfer, as the furnace temperature increases above 400 °C, radiation becomes more important and eventually dominates since the radiative heat flux scales by $\boldsymbol{q} \propto T^4$. The impact of radiation is complicated and competitive. On one hand, radiation from the inner walls of the furnace through the glass pellet could lead to a more uniform temperature distribution between the pellet and the furnace. On the other hand, radiation away from the TEM grid may lead to the evolution of a temperature discrepancy between the grid and the furnace thermocouple. This second mechanism was identified and explored by Mortensen and colleagues, who showed that a temperature discrepancy of ~5 °C evolved for furnace temperatures up to 700 °C [25]. The softening point of borosilicate glass is approximately 800 °C, so it is advised to operate the reactor below 700 °C to avoid damage to the glass fiber pellet-loaded furnace holder. The magnitude of this temperature difference is on the same order of that reported in Figure 5, and therefore, for specimen temperatures up to 700 °C, the temperature distribution in the reactor is unlikely to change by a significant amount compared with presented below.

## 3. Results and Discussion

Steady-state simulations were performed with a reactant gas inflow of 1 SCCM of stoichiometric CO and $O_2$, which leads to a chamber pressure of ~2.2 – 2.4 Torr. The furnace temperature was set in the range of 25 – 340 °C. These conditions correspond to those where experimental *operando* data were acquired, which serves as a basis for evaluating the simulations.

*3.1. Velocity and Pressure Profiles*

A view of the velocity field and pressure profiles around the *operando* pellet reactor are shown in Figure 3. Data from a temperature setpoint of 100 °C are shown at the top half of the figure, while data from 300 °C are shown at the bottom for comparison. The plots of the velocity fields contain overlaid quiver plots to show the bulk flow velocity vectors within the fluid regions of the cell; the velocity vectors are plotted for magnitudes greater than 3 m s$^{-1}$ and their lengths are displayed on a natural log scale. As seen in Figures 3a and 3c, the bulk fluid velocity is low everywhere in the cell except for in the immediate vicinity of the differential pumping aperture outlets. While the gas there accelerates rapidly from the cell- averaged speed of ~0.2 m s$^{-1}$ up to ~100 m s$^{-1}$, the velocity field only extends ~1 mm into the chamber and does not produce appreciable flow near the heater holder or *operando* pellet surfaces. The velocity fields at 100 °C and 300 °C show little difference. The gradient at 300 °C is slightly more confined to the aperture outlets due to the lower average cell pressure. The broad uniformity of low-magnitude velocities throughout the cell suggests that convective heat transfer and mass transport are limited (see dimensionless number analysis presented in Supplemental Appendix 3).

The pressure profiles in Figures 3b and 3d follow a trend similar to the velocity distributions. Far from the pumping aperture outlets, the pressure in the cell is uniform. Changes in pressure are only seen to occur in the immediate vicinity of the outlets, where a pressure drop forms to drive the bulk flow of gas out of the cell. This area of interest is denoted with a red arrow in the figure. From these results, one can justifiably believe that the steady-state pressure measured along the inlet line matches the pressure around the catalyst. This is a particularly useful finding for those interested in detecting *operando* conversions by accounting for changes in total cell pressure [61]. For the CO oxidation reaction, $CO + \frac{1}{2}O_2 \rightarrow CO_2$, the pressure in the cell decreases as 1½ mol of reactants are converted to 1 mol of products. The effect is visible as the pressure in the cell drops 0.2 Torr from the conversion of CO and $O_2$ to $CO_2$ when the furnace is heated from 100 °C (Figure 3b) to 300 °C (Figure 3d). Finally, it is noted that the velocity and pressure distributions simulated here match well with those reported by Mortensen and coworkers [25].

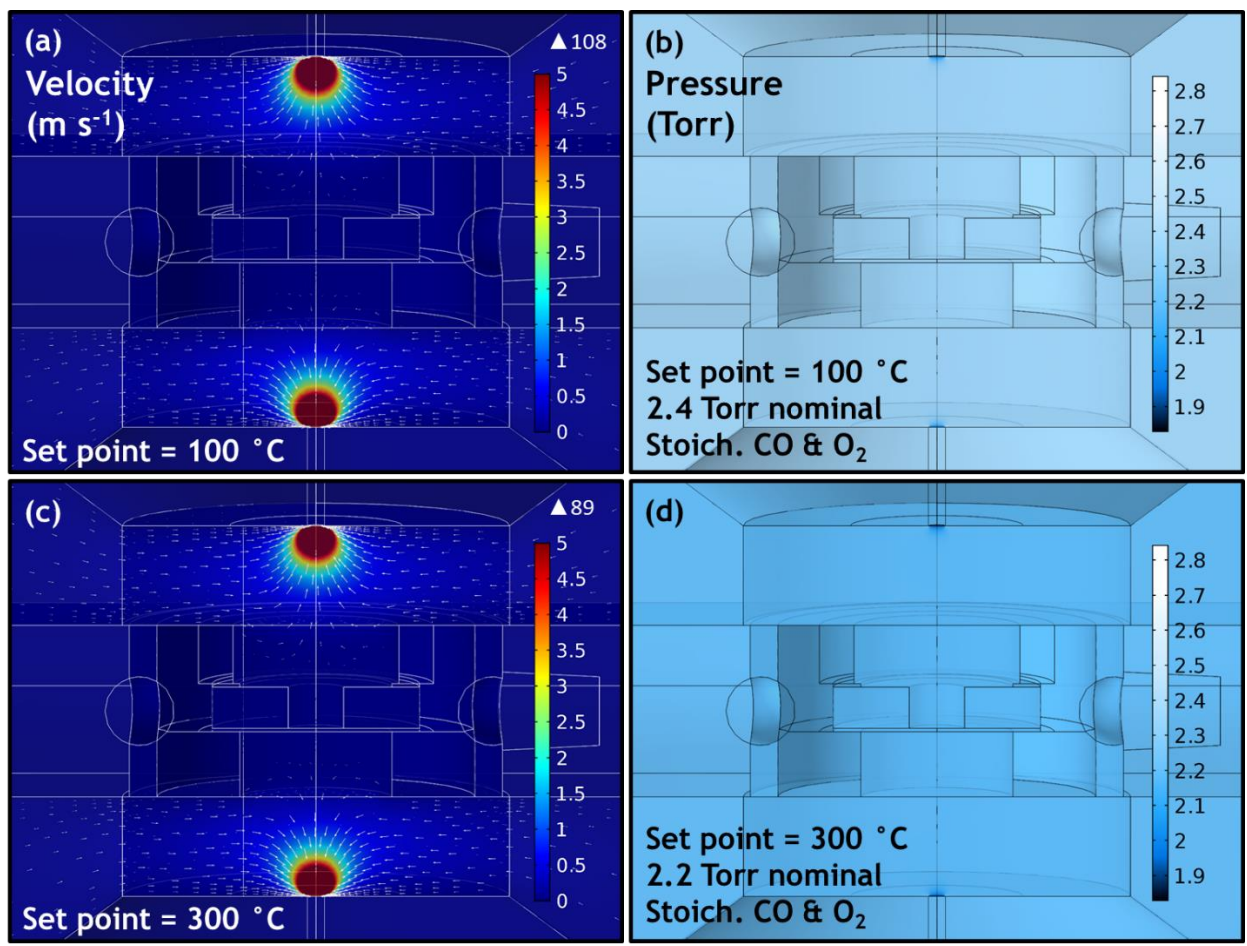

**Figure 3.** Velocity fields and pressure distributions near the *operando* pellet reactor and differential pumping aperture outlets for a temperature set point of **(a, b)** 100 °C and of **(c, d)** 300 °C. The velocity and pressure are both uniform in regions away from differential pumping aperture outlets, and so those regions are not shown here. The velocity vectors in **(a)** and **(c)** are plotted for magnitudes greater than 3 m s$^{-1}$ and their lengths are displayed on a natural log scale.

*3.2. Temperature Distribution*

Figure 4 displays the temperature distribution around the *operando* pellet reactor for four furnace thermocouple set points: (a) 100 °C, (b) 190 °C, (c) 230 °C, and (d) 300 °C. The elevated temperature is localized to the reactor and the gas surrounding it, so the discussion is focused on this region. For each condition, the apparent temperature of the reactor matches remarkably well with the furnace set point. The temperature distributions are also largely uniform throughout the reactor. The temperature rapidly decreases outside this region to 25 °C at the surface of the water-cooled pole pieces. The thermal gradient that exists between the furnace and the pole pieces deepens at higher temperatures but shows equivalent behavior.

The high degree of thermal uniformity in the *operando* pellet reactor is an important and encouraging result that suggests that the furnace temperature set point matches the temperature at the TEM grid and the *operando* pellet. A quantitative comparison of the temperature difference at these locations vs. the temperature set point is shown in Figure 5. The temperature of the thermocouple is described by a point value at that location, while the TEM grid and *operando* pellet temperatures are defined by averages over the respective surface and volume domains. A positive value signifies that the location is hotter than the temperature set point. At 25 °C all the temperatures overlap. For furnace set points below ~220 °C, the temperature of the grid and pellet are lower than the set point by 0.5 – 1 °C. This slight gradient drives the flow of heat to them from the hotter furnace. The smaller thermal conductivity of the glass pellet causes the average pellet temperature to be slightly (~0.5 °C) lower than the highly conductive tantalum TEM grid. The inflection point observed in the curves at ~190 °C indicates the ignition of the exothermic catalytic reaction. At set points above ~250 °C, the contribution from the heat of the reaction exceeds the heat lost to the surrounding gas, and the pellet/grid become ~1 °C hotter than the furnace itself.

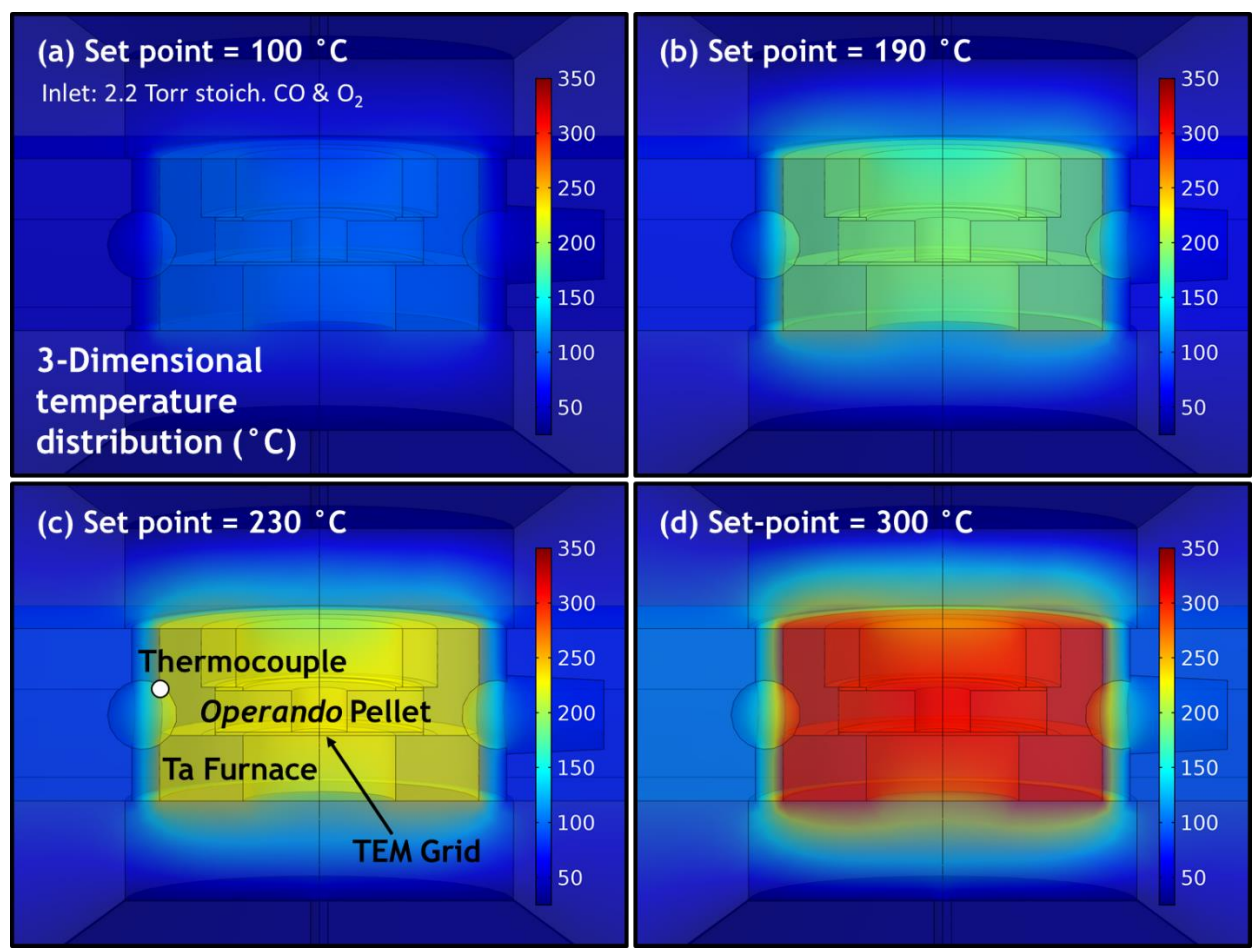

**Figure 4.** 3-dimensional temperature distributions in and around the *operando* pellet reactor during catalysis for four furnace thermocouple set points: **(a)** 100 °C, **(b)** 190 °C, **(c)** 230 °C, and **(d)** 300 °C. At each condition the temperature distribution appears largely uniform and matches well with the furnace thermocouple set point.

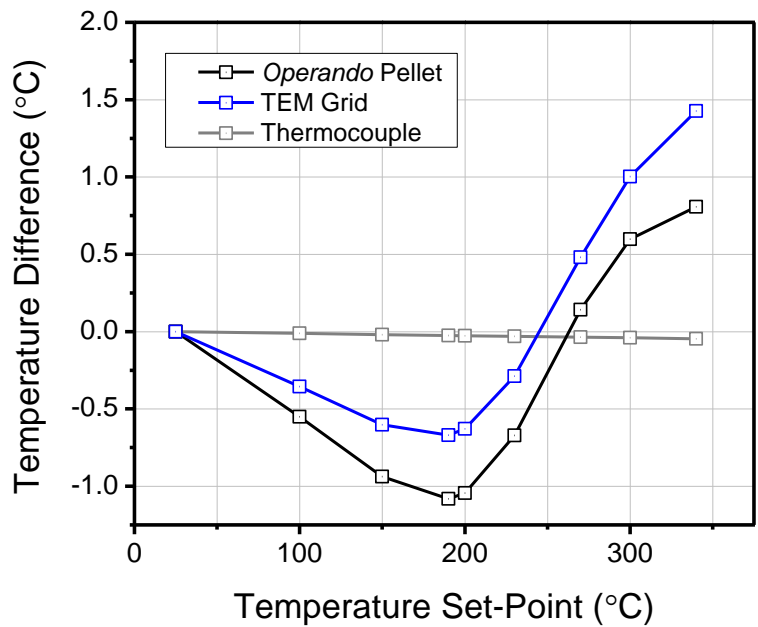

**Figure 5.** Quantitative comparison of the temperature set point vs. the simulated temperature at the furnace thermocouple (gray line), TEM grid (blue), and *operando* pellet (black). A positive value signifies that the location is hotter than the temperature set point. The inflection point at ~190 °C signifies the ignition of the catalytic reaction.

The temperature difference at the thermocouple is virtually 0 °C for all set points. It is worth noting that an absence of a temperature difference at the thermocouple vs. the furnace set point matches the simulations done by Mortensen and coworkers [25]. Overall, for the conditions simulated here, the TEM grid temperature does not differ from the intended furnace set point by more than 1.5 °C. This data suggests that the furnace thermocouple is a reliable probe of the temperature of the catalyst imaged on the TEM grid – even at temperatures where the catalyst is active.

### 3.3. Distribution of Catalytically Produced $CO_2$

The steady-state 3-dimensional mole fraction of catalytically produced $CO_2$ is shown in Figure 6 for thermocouple set points of (a) 100 °C, (b) 190 °C, (c) 230 °C, and (d) 300 °C. Here, the mole fraction of species $i$ is defined in a standard way as the number of moles of that species divided by the total number of moles of all species within the region of consideration (see, for example, Supplemental Equation S21). The gas composition is homogeneous in the cell except for the region in and around the *operando* pellet reactor, so the results and discussion presented below focus on this region. The inset of each figure displays the simulated CO conversion, $X_{CO}$, at each temperature. Here, $X_{CO}$ is defined as:

(15) $\quad X_{CO} = \frac{\dot{n}_{CO,in} - \dot{n}_{CO,out}}{\dot{n}_{CO,in}}$

Where $\dot{n}_{CO,in}$ is the molar flow rate (mol/s) of CO into the ETEM chamber, which is calculated in the model by taking a surface integral of the molar flux of CO across the inlet surface into the ETEM. The variable $\dot{n}_{CO,out}$ is the molar flow rate of CO out of the ETEM, which is calculated similarly by taking a surface integral of the molar flux of CO across the differential pumping aperture outlet surfaces.

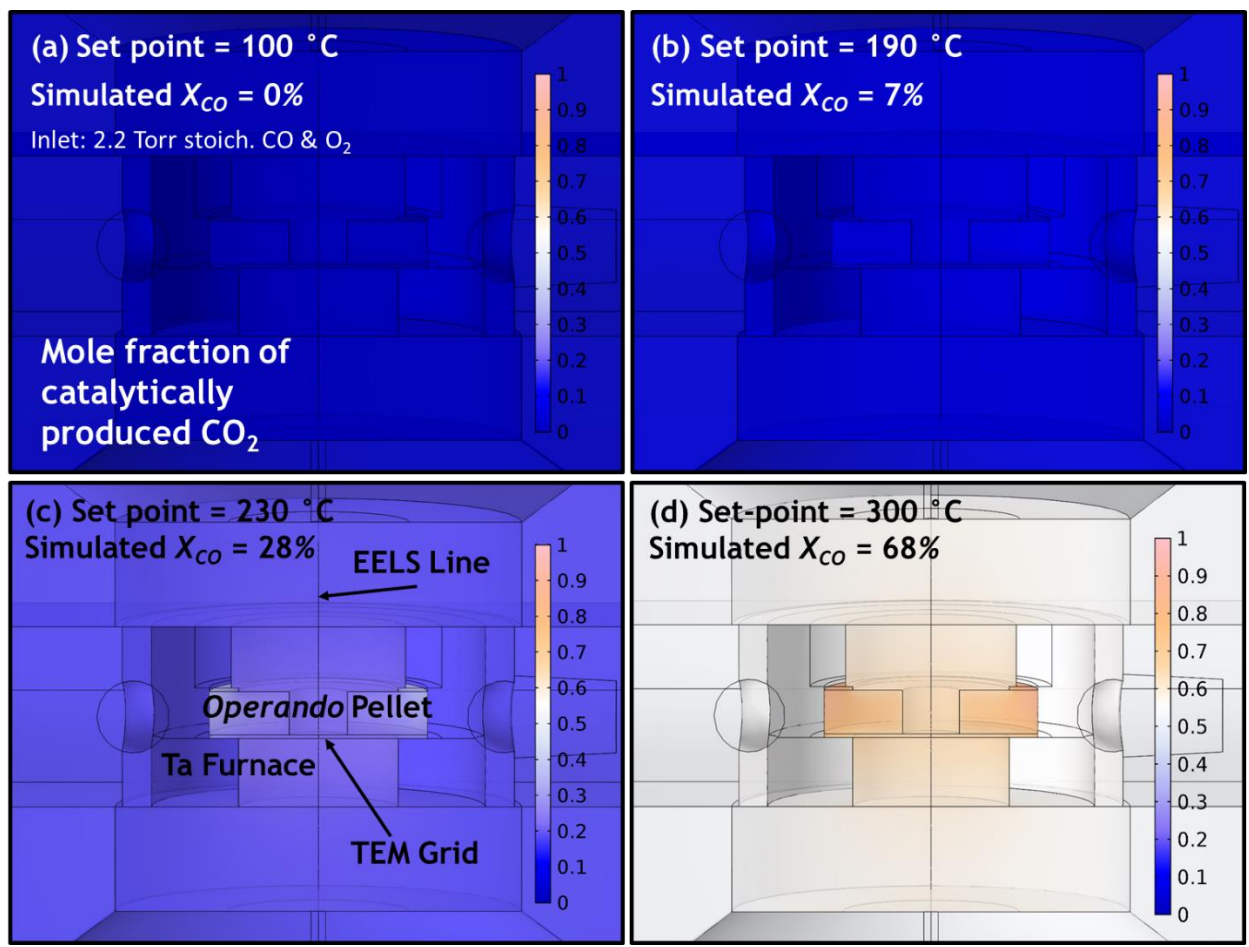

**Figure 6.** Mole fraction of catalytically produced $CO_2$ in and around the *operando* pellet reactor for four furnace thermocouple set points: **(a)** 100 °C, **(b)** 190 °C, **(c)** 230 °C, and **(d)** 300 °C. The simulated CO conversion, $X_{CO}$, is given at each condition in the inset of the respective figure.

Below the catalyst's light off point, very little $CO_2$ is produced, as observed in Figure 6a, which shows that the $CO_2$ mole fraction at 100 °C is approximately 0 everywhere. As the temperature ramps up the catalyst becomes active and the amount of $CO_2$ produced increases. The $CO_2$ produced from the reaction distributes mostly homogeneously throughout the cell and the reactor. Figure 6b shows the largely uniform $CO_2$ mole fraction at 190 °C, a condition of 7% conversion. As the conversion increases, a slight enrichment of product gas develops in the *operando* pellet, in particular at the top outermost region of the pellet, which is enclosed by the impermeable surfaces of the Ta furnace and washer. The bottom layer of the pellet is exposed to the TEM grid, which has been modeled as porous and thus accessible for $CO_2$ diffusion. Figure 6c and 6d present the 3D mole fraction of $CO_2$ at 230 °C and 300 °C, where the simulated conversions of CO are 28% and 68%, respectively. Even at such high catalytic conversions, the gas in and around the reactor exhibits a well-mixed composition.

A deeper understanding of the compositional variation throughout the reactor is developed by comparing the average $CO_2$ mole fraction quantified at three domains of interest across a range of temperatures. Figure 7a displays the $CO_2$ mole fraction averaged over the *operando* pellet (black line), at the TEM grid (blue line), and along the EELS line (red line), for furnace temperature set points spanning 100 – 340 °C. Experimental measurements of the $CO_2$ composition along the EELS line (red boxes) are also plotted from nominally identical conditions across the same range of temperatures. The experimental data agree well with the simulated EELS measurements. The high degree of agreement between the experimentally acquired and computationally modeled $CO_2$ compositions across a broad range of temperatures is taken as evidence that the model faithfully captures the relevant heat transport and chemical reaction physics. The catalyst is seen to begin lighting off at 190 °C. At each location, the $CO_2$ composition curve displays a sigmoidal shape. The average $CO_2$ mole fraction in the *operando* pellet is higher than that along the EELS line, though this is expected given the 3-dimensional distributions shown in Figure 6. The difference between the average $CO_2$ mole fraction in the pellet and that along the EELS line grows as the conversion increases, reaching a maximum of 0.15 at 300 °C, where the $CO_2$ mole fractions in the pellet and along the EELS line are simulated to be 0.75 and 0.60, respectively. The difference between the pellet and EELS line drops to 0.03 when the activity is lower at 190 °C. The percentage difference between the pellet and the EELS line is more constant across the same temperature range, as shown by the orange boxes in Figure 7b. At 190 °C the percent difference is 27%, and at 300 °C the percent difference is 20%.

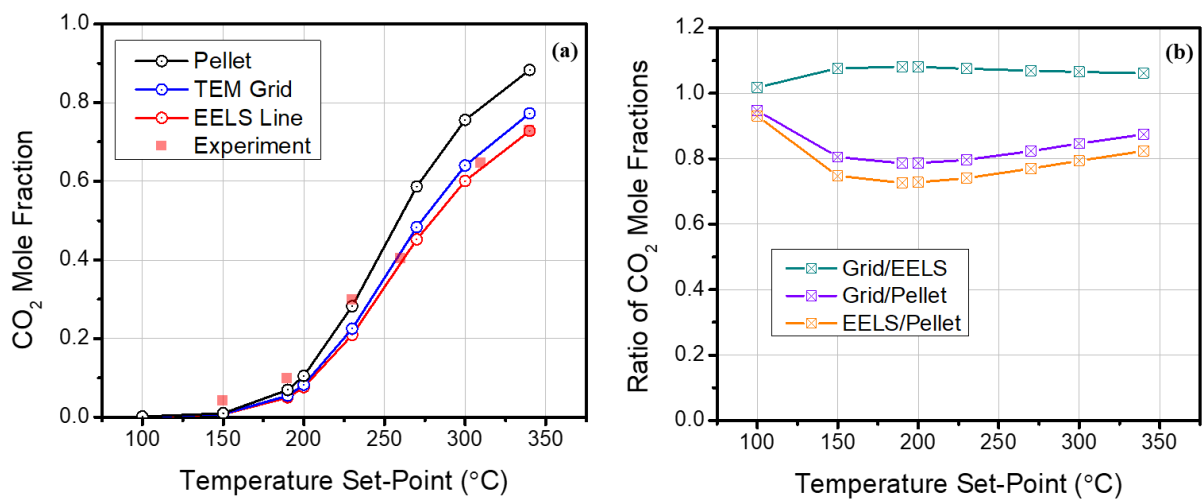

**Figure 7. (a)** Quantitative comparison of the $CO_2$ mole fraction across a range of temperatures in the *operando* pellet (black circles), at the TEM grid (blue circles), and along the EELS center line (red circles). The composition measured experimentally (red boxes) agrees very well with the values simulated along the EELS center line. **(b)** Ratio of $CO_2$ mole fraction between the grid and the EELS line (green boxes), the grid and the pellet (purple boxes), and the EELS line and the pellet (orange boxes).

While the difference in composition between the EELS line and the pellet can be appreciable, the difference between the EELS measurement and the composition at the TEM grid is rather small for all conditions. The composition at the TEM grid is important since the grid supports the catalyst that is actually imaged during an *operando* TEM experiment. In most cases here, the mole fraction difference is between 0.01 – 0.04, which is on the same order as the precision of the experimental gas quantification procedure [23]. As seen by the green line in Figure 7b, the fractional variation over the entire temperature range is 3 – 8%. Consequently, one can practically assume that the gas composition measured with EELS matches that surrounding the catalyst on the grid. This is an important result because it demonstrates that EELS can reliably probe the local gas composition around the imaged catalyst.

*3.4. CO Conversion Analysis*

Two important performance characteristics for studying catalytic reaction chemistry are the conversion of reactants and the rate of reaction. Here we show that EELS can be used to estimate both the CO conversion, $X_{CO}$, and the reaction rate, $r_{CO}$. Recall from Equation (15) that the CO conversion can be calculated from the integrated flux of CO across the gas inlet and pumping aperture outlet surfaces of the ETEM cell. The CO conversion obtained in this way is regarded as the true conversion in the chemical reaction engineering sense, and it is tabulated in the second column of Table 3 for reactor temperatures spanning 150 – 340 °C. The CO conversion can also be expressed in terms of the $CO_2$ mole fraction at the outlet surfaces. A derivation starting from Equation (15) is provided in Supplemental Appendix 8, and it results in the following expression for a stoichiometric mixture of reactants:

(16) $X_{CO} = \dfrac{y_{CO_2}}{(1-y_{CO_2})\times\frac{2}{3} + y_{CO_2}}$

Here, $y_{CO_2}$ is the $CO_2$ mole fraction of the gas leaving the ETEM reactor through the pumping aperture outlets. It is not feasible to measure the composition across this exit surface experimentally. However, given the large degree of gas-phase homogeneity, an estimate may be made from the $CO_2$ mole fraction measurement obtained by EELS. The third column of Table 3 shows the estimated CO conversion obtained in this way, *i.e.*, by calculating $X_{CO}$ in Equation (16) from the $CO_2$ mole fraction along the EELS line. Estimates of $X_{CO}$ are shown for furnace temperature set points spanning 150 – 340 °C.

**Table 3**

Comparison of CO conversion obtained by integrating the flux of CO in and out of cell, and by estimating the outlet composition from EELS. The relative difference between the two conversion values is also given.

| Temperature (°C) | $X_{CO}$ Calculated from CO Flow Rate in and out of ETEM Cell (%) | $X_{CO}$ Estimated from $CO_2$ Mole Fraction along EELS Line (%) | Relative Difference between Calculated and Estimated $X_{CO}$ (%) |
|---|---|---|---|
| 150 | 1.1 | 1.0 | 9.1 |
| 190 | 7.1 | 7.4 | 4.2 |
| 200 | 10.5 | 11.1 | 3.7 |
| 230 | 27.5 | 28.4 | 3.3 |
| 270 | 54.0 | 55.3 | 2.4 |
| 300 | 67.8 | 69.3 | 2.2 |
| 340 | 78.6 | 80.0 | 1.8 |

The CO conversion values calculated from the CO flow in and out of the ETEM cell differ little compared to those estimated from the gas composition along the EELS line. The relative difference for each temperature is provided alongside the conversions in Table 3. At 190 °C, the calculated and estimated CO conversions are 7.1% and 7.4%, respectively, which corresponds to an absolute difference of 0.003 and a relative difference of 4.2%. When the catalyst is more active at 270 °C, the calculated and estimated conversions are 54.0% and 55.3%, respectively, which corresponds to an absolute difference of 0.013 and a relative difference of 2.4%. The conversion estimated from EELS is consistently higher by a slight amount, due to the enrichment of $CO_2$ along the EELS line relative to the composition at the outlet (see, *e.g.*, Figure 6c or 6d, particularly the region along the EELS line in the vicinity of the pellet). Overall, though, the resulting impact on estimating the CO conversion is small, amounting to a relative difference around 2 – 9% for the conditions explored here. As a result, these findings show that EELS can be used to estimate the catalytic conversion during an open-cell *operando* TEM experiment. Moreover, the model developed here may be used to provide a correction factor to obtain the true conversion if desired.

These simulations along with experimental measurements show that the gas is well-mixed in differentially pumped cells [22]. This relative homogeneity for the differentially pumped cell combined with the *operando* pellet architecture means that to a good approximation, the system behaves as continuously-stirred tank reactor (CSTR), especially at low conversions. The ability to apply a simple reactor model to the *operando* pellet reactor would greatly facilitate the evaluation of kinetic parameters for catalytic structure-activity relationships (*e.g.*, activation energies), which is a subject of future work for this project.

*3.5. Reaction Rate Analysis*

With an estimate of the CO conversion, one can calculate the rate of product formation, or the reaction rate, which is of principle importance to chemical kinetics. In the model, the true rate of product formation (with units of mol $CO_2$ per second) may be found by integrating the reaction rate throughout the domain of the pellet where the reaction occurs. For heterogeneous catalysis, the reaction rate is usually normalized to the mass of catalyst or to the surface area of the catalyst in the reactor. With an appropriate active site model, one may also normalize the rate to the number of active sites available, yielding a turnover frequency measurement. Here we have chosen to normalize to mass, with the total mass of catalyst loaded in the pellet as 200 µg, since this is the amount loaded during a typical *operando* experiment with the Ru/$SiO_2$ catalyst studied here [62]. The mass-normalized rate integrated over the entire domain of the pellet, which we refer to as $r_1$, with units of mol $CO_2$ per second per gram catalyst, is tabulated across a range of temperatures in Table 4. The CO conversion estimated from EELS is provided again in this table for reference.

An estimate of integrated rate of product formation may be made from the conversion measurement derived from EELS and is simply related to the estimated $X_{CO}$ and to the inlet flow rate of reactants. For reference, the inlet molar flow rate of CO into the cell, $\dot{n}_{CO,in}$, is ⅔ SCCM or $4.95 \times 10^{-7}$ mol/s. The estimated rate of product formation, $r_2$, may be calculated and normalized to the mass of catalyst in the pellet, $m_{cat}$, by the equation that follows:

(17) $r_2 = \frac{X_{CO} \times \dot{n}_{CO,in}}{m_{cat}}$

The fourth column of Table 4 displays the mass-normalized rate estimated by the EELS conversion measurement. The relative difference between the rate integrated over the entire domain of the pellet ($r_1$) and that estimated by the EELS conversion measurement ($r_2$) is also given in Table 4.

It is of interest to compare the estimated – and experimentally measurable – rate of product formation to the true rate obtained through 3D integration. As seen in Table 4, in general, for reactant conversions above 5%, the EELS measurement provides an estimate of the rate of product formation that is within 10% of the true, integrated rate. Typically, the systematic error in the EELS conversion is 5 – 7% larger than the true conversion. Consequently, the error on the ratio of rates at different temperatures is generally better than 5%. Pragmatically, the inlet reactant flow rates are calibrated and known, so these findings present an important result demonstrating that one can use EELS to determine the overall steady-state reaction rate during an open-cell *operando* TEM experiment. If desired, simulations with the model developed here could be performed to obtain a correction factor that adjusts experimental measurements to the true rate in the simulation. Finally, it is noted that the integrated reaction rate (moles of CO consumed per second) approaches the rate of CO flown into the cell. The only way this could be true is if most (≥ 80%) of the reactant gas interacts with the catalyst-loaded pellet before exiting through the pumping aperture outlets.

The overall rate of product formation discussed above and estimated by Equation (17) is averaged over the entire pellet in which the mass distribution and reactant concentrations vary (especially at high conversions). We can determine a local reaction rate by noting that the rate depends on the local concentration of reactants (see Equation (14)). Of particular interest is the reaction rate of the catalytic particles on the TEM grid, since those are the particles that are imaged. Figure 6 shows that the gas composition on the TEM grid and on surface along the inner hole in the pellet are almost identical. We can therefore calculate the reaction rate at the surface along the inner hole in the pellet to determine the reaction rate of the catalytic particles on the grid. We determine the reaction rate per unit mass by integrating the rate and mass around a 50 μm thick layer at the surface of the inner hole in the pellet. This reaction rate, which we refer to as $r_3$, is shown as a function of temperature in Table 4. The ratio of the reaction rate at the surface of the pellet ($r_3$) to the reaction rate integrated over the whole pellet ($r_1$) is also displayed in the table. In general, the reaction rate averaged over the entire pellet is lower than that at the TEM grid because the concentrations of reactants in the pellet are lower than at the grid (due to mass transport limitations). The difference between the two rates increases with conversion. For conversions below 30%, the reaction rate for catalyst particles on the grid is less than 15% higher than that of the average rate in the pellet. However, at a conversion of 80%, the reaction rate on the grid is more than a factor of two higher than the rate integrated over the entire pellet.

**Table 4**
Summary of reaction rate analysis. The rate of $CO_2$ formation obtained by integrating the reaction rate over the entire pellet ($r_1$) and the rate of $CO_2$ formation estimated by the EELS CO conversion measurement ($r_2$) are tabulated as a function of temperature. The ratio between the integrated and estimated rate is given ($r_2:r_1$), and the CO conversion estimated by EELS is provided again for reference. Finally, the rate of $CO_2$ formation obtained by integrating the reaction rate within a 50 μm thick layer at the surface along the inner hole in the pellet ($r_3$) is tabulated, in addition to the ratio between the surface-integrated and pellet-integrated rates ($r_3:r_1$).

| Temperature (°C) | $X_{CO}$ Estimated from EELS (%) | $r_1$, Rate Integrated over Entire Pellet (mol $CO_2$ sec$^{-1}$ $g_{cat}^{-1}$) | $r_2$, Rate Estimated by EELS Conversion (mol $CO_2$ sec$^{-1}$ $g_{cat}^{-1}$) | Ratio of $r_2:r_1$ (%) | $r_3$, Rate at Surface of Inner Hole of Pellet (mol $CO_2$ sec$^{-1}$ $g_{cat}^{-1}$) | Ratio of $r_3:r_1$ (%) |
|---|---|---|---|---|---|---|
| 150 | 1.0 | $2.19 \times 10^{-5}$ | $2.45 \times 10^{-5}$ | 111.9 | $2.14 \times 10^{-5}$ | 97.7 |
| 190 | 7.4 | $1.65 \times 10^{-4}$ | $1.81 \times 10^{-4}$ | 109.7 | $1.66 \times 10^{-4}$ | 100.6 |
| 200 | 11.1 | $2.49 \times 10^{-4}$ | $2.71 \times 10^{-4}$ | 108.8 | $2.56 \times 10^{-4}$ | 102.8 |
| 230 | 28.4 | $6.49 \times 10^{-4}$ | $6.97 \times 10^{-4}$ | 107.4 | $7.40 \times 10^{-4}$ | 114.0 |
| 270 | 55.3 | $1.27 \times 10^{-3}$ | $1.36 \times 10^{-3}$ | 107.1 | $1.83 \times 10^{-3}$ | 144.1 |
| 300 | 69.3 | $1.60 \times 10^{-3}$ | $1.70 \times 10^{-3}$ | 106.3 | $2.80 \times 10^{-3}$ | 175.0 |
| 340 | 80.0 | $1.85 \times 10^{-3}$ | $1.96 \times 10^{-3}$ | 105.9 | $4.09 \times 10^{-3}$ | 221.1 |

The current model may overestimate the difference in reaction rates since we essentially assume that the pellet is sealed when it contacts the body of the hot stage, which limits mass transport and allows the $CO_2$ concentration to build up (see Figure 6). In practice, the pellet will not make a gas-tight seal with the body of the hot stage, and consequently the difference between the reaction rate in the pellet and at the TEM grid will be less. However, for the current reaction and catalyst, the model suggests that for conversions less than 50%, the absolute value of the reaction rate can be determined to within about 20%. The model can be used to make more accurate estimates of this difference provided the catalyst loading is known and the order of the reaction kinetics is known. Of course, relative differences in reaction kinetics will be known more precisely.

The ability to measure the catalyzed reaction rate provides useful information for catalyst chemistry and characterization. Generally speaking, a measurement of the reaction rate allows an experimentalist to directly correlate atomic-level imaging and spectroscopy with the chemical kinetics of the same catalyst. For the ETEM reactor, at low conversions the relative homogeneity in the gas composition within the domain of the TEM pellet and grid enables reaction rates to be estimated from EELS for the catalyst particles being imaged. Even at higher conversions, the entire TEM grid used for imaging is exposed to the same reactor conditions and surface chemistry taking place at similar structures on different nanoparticles throughout the TEM sample will be identical. Essentially, for the differentially pumped ETEM reactor, the reactor conditions are fairly uniform, well defined, and measurable. The *in situ* TEM approach described here can be called *operando* because the observed nanoparticle structure can be directly linked to chemical kinetics. This *operando* capability may permit one to differentiate catalytically-relevant structures from spectator species by identifying those which emerge or correlate with the activity of the catalyst.

## 4. Conclusions

We have developed a finite element model combining fluid dynamics, heat transfer, multi-component mass transport, and chemical reaction engineering in order to determine the gas and temperature profiles present during catalysis in an *operando* experiment performed in an open cell ETEM. The model determines steady state solutions for an ETEM reactor with inflows of reactant gas mixtures, with product gases produced in the microscope during catalysis. Under typical *operando* TEM conditions, mass transport is dominated by diffusion, while heat transfer is dominated by conduction. For reactor temperatures above 400 °C, radiation becomes more important and eventually dominates heat transfer.

As a case study, we have applied the model to a SiO$_2$-supported Ru catalyst performing CO oxidation. Steady state solutions were computed for a 3 mbar reactant gas inflow of stoichiometric CO and O$_2$. The simulated composition of catalytically-produced CO$_2$ agrees well with experimental measurements taken under nominally identical conditions across a range of temperatures spanning 25 – 350 °C. The CO$_2$ produced from the reaction distributes throughout the cell and the reactor, with an enrichment in the *operando* pellet. The enrichment in the *operando* pellet relative to the TEM grid ranges from 21% to 12% for CO$_2$ mole fractions between 5% and 73%, as measured with EELS, respectively. The gas composition at the TEM grid, which is important as the grid contains the catalyst that is imaged during an experiment, differs by less than 8% from the composition measured with EELS. For the conditions simulated here, the average temperature at the TEM grid differs from the intended furnace set point by less than 2 °C, even at temperatures where the catalyst is active. The results show that one can determine the temperature and gas composition surrounding catalytic nanoparticles imaged during an *operando* experiment in a differentially-pumped ETEM.

In general, the simulations show that the temperature and gas are relatively homogeneous within the hot zone of the holder where the catalyst is located. The uniformity of gas and temperature across the catalyst and TEM sample indicates that the system behavior around the catalyst approximates that of a continuously stirred tank reactor. The results show that EELS can be used to estimate the catalytic conversion of reactants in the ETEM cell to within 10%. A kinetic analysis shows that the rate at which reactants are consumed in the pellet approaches the rate at which reactants are flown into the ETEM cell, which demonstrates that most of the reactant gases interact with the catalyst-loaded pellet and that there is limited gas bypass. A very important consequence of the gas-phase homogeneity is that the overall activity at all points in the TEM catalysts are similar to within 20% for conversions below 50%. Essentially, the reactor conditions are fairly uniform, well defined, and measurable for a differentially-pumped ETEM reactor. It is this characteristic that allows one to claim that this is truly *operando* TEM since the observed nanoparticle structure can be directly linked to known local reactor conditions and chemical kinetics. Overall, these findings indicate that under suitable conditions during an *operando* ETEM experiment, one can reliably evaluate the temperature and steady-state reaction rate of the catalyst that is imaged on the TEM grid.

## Acknowledgements

This work was supported by the National Science Foundation (NSF) and Arizona State University. NSF grant CBET 1604971 supported JAV, JLV, and PAC who refined and tested the finite element model. JLV and PAC wrote the manuscript. NSF grant CBET-1134464 supported BJM, JTL, and PAC who acquired the experimental data and started the development of the finite element model. We thank Arizona State University's John M. Cowley Center for High Resolution Electron Microscopy for microscope use.

# Supplementary Information

Chemical Kinetics for *Operando* Electron Microscopy of Catalysts: 3D Modeling of Gas and Temperature Distributions During Catalytic Reactions

Joshua L. Vincent[1], Jarod W. Vance[1], Jayse T. Langdon[1,2], Benjamin K. Miller[1,3], and Peter A. Crozier[1]*

[1]*School for Engineering of Matter, Transport, and Energy, Arizona State University, Tempe, Arizona 85281*

[2]*Present address: Department of Mechanical Engineering, University of Texas at Austin, Austin, Texas 78712*

[3]*Present address: Gatan, Inc., Pleasanton, CA, USA.*

*Corresponding author email: crozier@asu.edu

# Extended view of ETEM model geometry

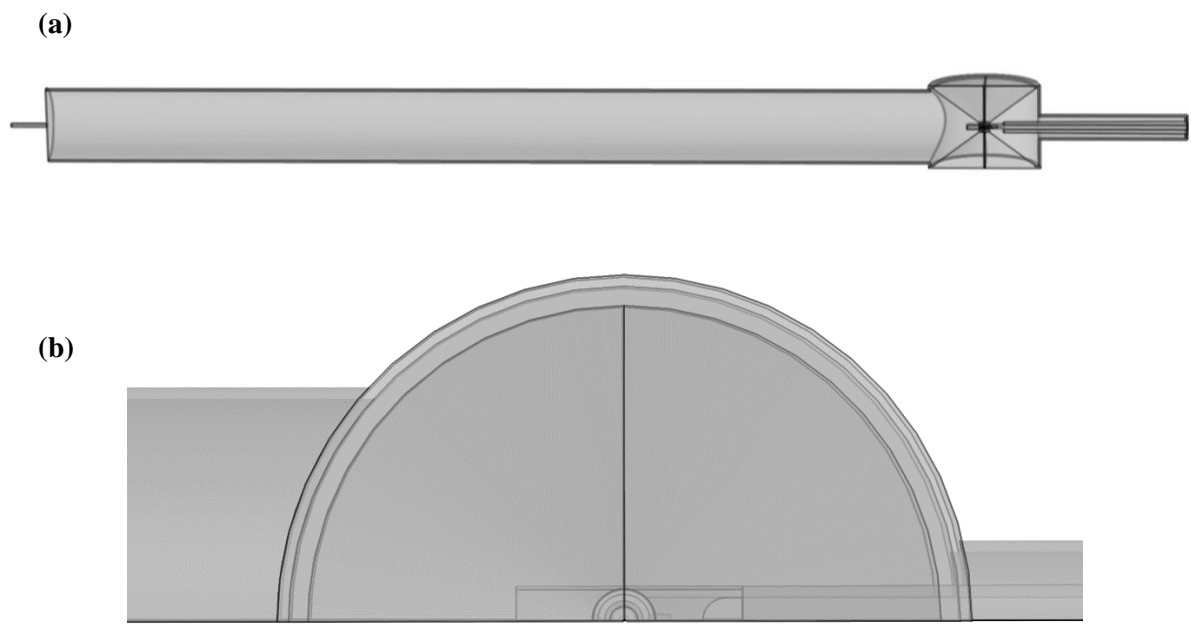

**Figure S1. (a)** Full view of model geometry showing reactant gas inlet into ETEM cell which extends approximately one meter out of the cell so that the gas composition in the cell does not impact the inlet composition. **(b)** Top-down view of the model geometry focused on the ETEM cell and *operando* pellet reactor, which makes both the cylindrical nature of the chamber and the planar symmetry of the model more apparent.

# Appendix 1: Geometry sensitivity analysis

Since some geometrical dimensions in the cell were not known exactly, a sensitivity analysis was performed to investigate the effect of small changes in the geometry on the results. The furnace heater holder and pellet reactor were directly examinable and their dimensions measurable, so no sensitivity analysis was performed on its dimensions. To perform the sensitivity analysis the dimensions of various geometrical components of the cell were changed by up to $\pm 15\%$. The dimensions of the pole piece gap, the width of the pole piece, and the width of the differential pumping aperture outlets were varied. Simulations were performed for an inlet reactant gas mixture of ~2.2 Torr of stoichiometric CO and $O_2$ with a thermocouple set point of 230 °C. Note that this corresponds to a condition when the catalyst is active and producing $CO_2$. The effect of geometric variations on the temperature, pressure, and $CO_2$ mole fraction of the gas located in the domain in the middle of the furnace holder was investigated. The results for this location are reported as it is of primary interest to the accuracy of the model. The results are plotted in Figure S2 below as a function of the parameter fraction, where a parameter fraction of 1.0 represents a geometric dimension that has been unchanged, a value of 0.9 represents a dimension that has been diminished by 10%, etc. A summary of the real values of the dimensions for each parameter fraction is provided in Table S1.

**Table S1.** Dimensions explored in the geometry sensitivity analysis.

| Pole Piece Gap | | Pole Piece Lens Width | | Aperture Outlet Diameter | |
|---|---|---|---|---|---|
| Parameter Fraction | Dimension (mm) | Parameter Fraction | Dimension (mm) | Parameter Fraction | Dimension (μm) |
| 0.9 | 4.86 | 0.85 | 4.76 | 0.9 | 225 |
| 1.0 | 5.40 | 1.00 | 5.60 | 1.0 | 250 |
| 1.1 | 5.94 | 1.15 | 6.44 | 1.1 | 275 |

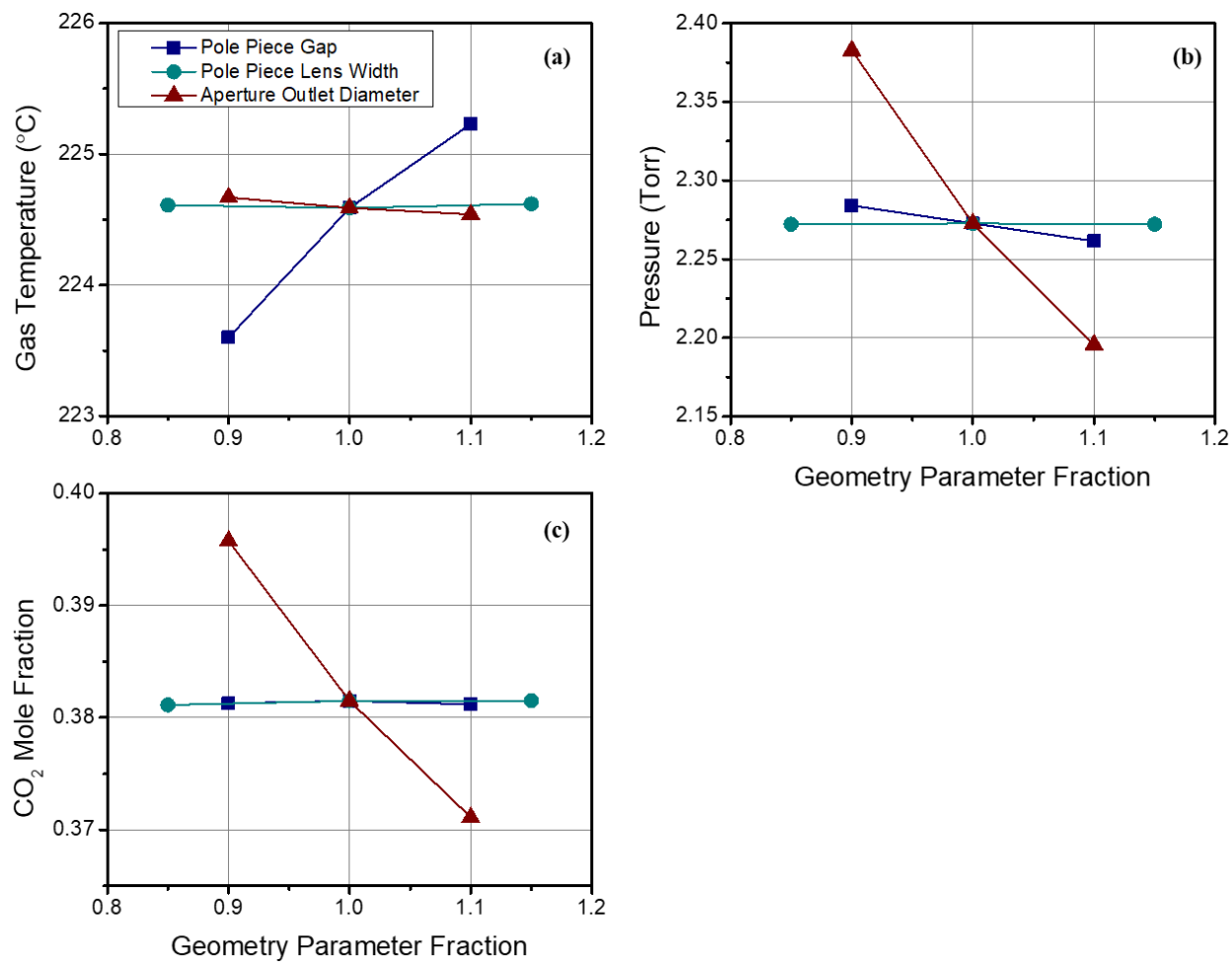

**Figure S2.** Effect of varying the dimensions of the pole piece gap (navy blue squares), the lens width (dark cyan circles), and the aperture outlet diameter (burgundy red triangles) on the **(a)** temperature, **(b)** pressure, and **(c)** $CO_2$ mole fraction of the gas located in the furnace holder.

Changing the dimensions of the pole piece lens width (dark cyan circles) by up to ±15% had a negligible impact on the results of the simulation. Varying the dimension of the pole piece gap (navy blue squares), which sets the distance between the hot furnace and the water-cooled pole pieces, had a slight impact on the temperature of the gas: when the gap dimension was decreased by 10% to 4.86 mm, the temperature of the gas dropped by 0.99 °C (since the hot gas can reach the cool pole pieces faster, thereby cooling it more), and when the gap dimension was increased by 10% to 5.94 mm, the temperature of the gas increased by 0.64 °C, or 0.28%. Varying the dimension of the pole piece gap changed the pressure slightly as well: when the dimension of the pole piece gap was reduced from 5.40 mm to 4.86 mm, the pressure of the gas increased from 2.27 to 2.28 Torr, and when the pole piece gap was increased from to 5.94 mm, the pressure decreased to 2.26 Torr. The trend can be understood by geometric arguments; as the space between the gap shrinks, the gas becomes more confined, and the pressure increases (by 0.01 Torr, or 0.4%). The converse is true as the gap increases. Alterations to the size of the gap had virtually no effect on the $CO_2$ mole fraction, changing it by less than 0.05 %.

Varying the diameter of the differential pumping aperture outlets altered the gas temperature by about only 0.05%. Changes to this dimension had a bigger impact than the others on the gas pressure and $CO_2$ mole fraction. This outcome is sensible as the aperture outlets serve as the only exit from the environmental cell. Increasing the outlet diameter from 250 to 275 μm thus allows more gas to leave the cell, which decreases the pressure from 2.27 to 2.19 Torr (by 0.08 Torr, or 3.5%). Increasing the aperture outlet diameter also reduces the $CO_2$ mole fraction, as the gas in that domain becomes diluted by pure CO and $O_2$ now coming from the inlet. As the aperture outlet diameter increases from 250 to 275 μm, the $CO_2$ mole fraction decreases from 0.381 to 0.371 (by 0.01, or 2.6%). It is worth noting that a variation of this magnitude is approximately the precision of the techniques used to measure the $CO_2$ mole fraction experimentally. Decreasing the diameter from 250 to 225 μm has a similar effect on increasing the pressure and $CO_2$ mole fraction.

Overall, this sensitivity analysis suggests that misestimations of up to ±15% on various geometric dimensions of the pole pieces would not significantly impact the behavior of the model or interpretation of its results. The geometry as-reported in the main text was thus used for all simulations.

## Appendix 2: Analysis of mesh quality

An analysis was performed to evaluate the quality of the mesh. The simulated conditions used in the calculations involve an inlet reactant gas mixture of ~2.2 Torr of stoichiometric CO and $O_2$ with a thermocouple set point of 230 °C. Note that this corresponds to a condition when the catalyst is active and producing $CO_2$. The quality of the mesh was varied from 105,000 to 766,000 elements while the temperature and $CO_2$ mole fraction of the gas located in the domain in the middle of the furnace holder was investigated. The pressure of the gas was observed to remain largely uninfluenced by changing the mesh quality, so it is not reported here. The results of the mesh quality analysis are shown below in Figure S3.

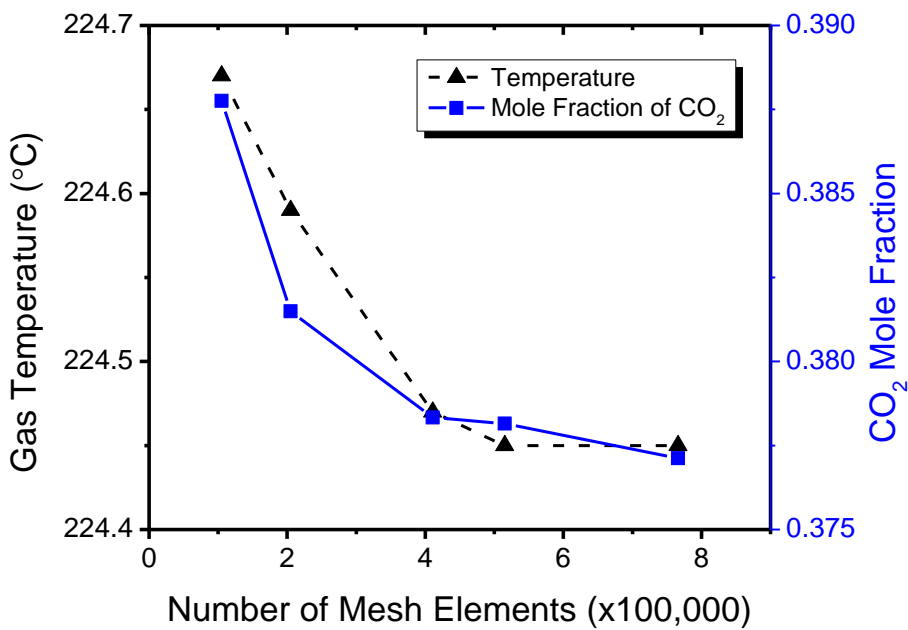

**Figure S3.** Effect of mesh quality in terms of number of elements on the temperature (black triangles, dashed line) and $CO_2$ mole fraction (blue squares, solid line) for gas located in the middle of the furnace holder.

The temperature is largely unaffected by the change in mesh quality, varying by < 0.3 °C as the number of mesh elements is increased from 105,000 to 766,000. The mole fraction of catalytically-produced $CO_2$ was slightly more sensitive to changes in the mesh quality, as it varied from 0.387 to 0.377 from the coarsest to the finest mesh. For a mesh containing 205,000 elements, the value was 0.381. As the number of mesh elements increased to $\geq 400{,}000$, the value stabilized to around 0.377. However, the computation time also increased rapidly. In typical simulations, numerous calculations were performed across a range of temperatures and usually across a range of another parameter space of interest, so a typical simulation required a number of hours to solve. Computing the solution to a model containing 400,000 elements required more than twice as much time (*i.e.,* many more hours) than one containing 205,000 elements. Given the marginal change in the computed $CO_2$ mole fraction, and the costly increase in computation time, a mesh containing 205,000 elements was selected to balance simulation speed and solution accuracy.

# Appendix 3: Dimensionless number analysis

## Mach number and numerical stability

A dimensionless number analysis was done to justify certain simplifications to the mass and heat transport equations solved by the model. First, to maintain numerical stability, the form of the Navier-stokes equations used in this model requires that the Mach number, $Ma$, stay below $Ma = 0.3$. The Mach number is calculated by Equation (S1) below:

(S1) $\quad Ma = \dfrac{v}{a}$

Here $a$ is the speed of sound in the gas and $v$ is the linear velocity. Given that the speed of sound in gas is generally greater than 200 m s$^{-1}$, and that the typical gas velocities in the simulation are low around 0.2 m s$^{-1}$, we expect that the Mach number will stay well below $Ma = 0.3$ and that the numerical stability requirement be satisfied.

## Reynolds number and laminar flow

The Reynolds number, $Re$, was computed to ascertain the flow regime of the gas in the ETEM cell, which is either laminar or turbulent. The Reynolds number is calculated by Equation (S2) below:

(S2) $\quad Re = \dfrac{\rho v L}{\mu}$

Here $\rho$ is the gas density, $L$ is a characteristic length, and $\mu$ is the gas viscosity. It is generally accepted that laminar flow occurs for $Re < 100 - 2{,}000$[1]. An upper bound on the Reynolds number can be computed by considering the lowest viscosity gas at its highest density and flow rate as it flows through the inlet. Take for example $O_2$ gas with a room temperature viscosity of $\mu = 2.08 \times 10^{-5}$ Pa*s at low pressure (see main text, Equation (5)). In this model the highest density is achieved at 4 Torr which provides a gaseous density of 6.83 g*m$^{-3}$ and an inlet linear gas flow rate of 0.239 m s⁻¹. The inlet diameter is taken to be the characteristic length and is equal to 0.026 m. Overall this produces a Reynolds number of $Re = 4.1 \ll 2{,}000$. Even near the differential pumping aperture outlets, where the simulated flow is calculated to be ~100 times higher, the Reynolds number would still lie well into the laminar flow regime. Therefore, all flow is treated as laminar in this model.

*Rayleigh number and natural convection*

The gravitational force in the Navier-Stokes equation was ignored on the basis that its primary effect, natural convection, was negligible. This simplification is supported by computing the Rayleigh number, $Ra$, which describes the significance of buoyancy-driven flow. The calculation for $Ra$ is given by Equation (S3) below:

(S3) $\quad Ra = \dfrac{g\beta \Delta T L^3}{\nu \kappa}$

Here $g$ is the standard gravitational acceleration, $\beta$ is the thermal expansion coefficient, $\Delta T$ is the temperature difference between the surface and the quiescent fluid, $L$ is a characteristic length, $\nu$ is the kinematic viscosity, and $\kappa$ is the thermal diffusivity. It is generally accepted that natural convection is insignificant for $Ra < 1{,}000\text{–}1{,}700$[2]. An upper bound for the Rayleigh number in the model can be computed by considering $CO_2$ which has a relatively low kinematic viscosity and thermal diffusivity. Consider a maximum temperature difference of $\Delta T = 400$ °C and take the cell height for the largest possible dimension, which gives $L = 0.04$ m. The value of $g$ is known to be 9.806 m s⁻². The thermal expansion coefficient of an ideal gas is known to be $\beta = \dfrac{1}{T}$, so for $T = 298$ K, we have that $\beta = 3.35 \times 10^{-3}$ K⁻¹. The kinematic viscosity and thermal diffusivity of $CO_2$ at 4 Torr and 298 K can be calculated from Equations (3), (6), (24), and (27) in the main text. These computations yield $\nu = 1.59 \times 10^{-3}$ m² s⁻¹ and $\kappa = 2.10 \times 10^{-3}$ m² s⁻¹. Overall this yields a Rayleigh number of $Ra = 251.1 < 1,000$. Considering the thermal profiles presented in the text, a more appropriate and typical Rayleigh number would consider the dimensions between the hot furnace, where the high-temperature gas is localized, and the water-cooled pole pieces. This dimension is 0.145 cm, which yields a Rayleigh number of $Ra = 0.01 \ll 1,000$. In both cases the dimensionless number analysis suggests that natural convection is not significant. Simulations that include a gravity force and permit natural convection were observed to be essentially identical to those without, so no gravity force is included in the model.

**Appendix 4: Experimental data for modeling flow out of pole piece apertures**

The flow of gas out of the ETEM cell is governed by effusion through the 250 μm differential pumping apertures. Rather than simulate this process explicitly in the model, experimental data was acquired and used to describe the static pressure of gas in the cell that resulted from a given inlet gas flow rate and composition. The variety of valve settings available in the ETEM vacuum system can lead to a number of possible cell pressures for a given fixed inlet flow rate of gas. Thus, specifying the particular configuration used is important when reporting a measurement of the static pressure that results for a given inlet flow rate of gas. For the data reported here, which was collected on an FEI Titan ETEM, the vacuum system was set to the E-TEM vacuum state. A known flow rate and composition of gas was flown into the column through valve *Vg1/2/3*, with the corresponding leak valve *LVg1/2/3* completely open. Valve *Vgi* was open, valve *Vrga1* was closed, and valve *Vrga2* was open. The leak valve *LVrga* was typically set to a value of 31,000, rendering it partially open. At the ETEM cell, the pumping valves *Vp0A/B/C* were all set to close, forcing the gas to be pumped out through the differential pumping apertures. The pressure in the cell was measured with the Pirani gauge (*PP/O*) or Baratron capacitance manometer (*BC/O*) located adjacent to the environmental cell. This configuration was used to acquire all of the data presented below and discussed in the main text.

The empirical data was fit with a parabolic function, which provided the pressure-flow relationship for the model. The empirical data for CO and $O_2$ and fitted parabolas are plotted below in Figure S4a. The behavior for mixtures was described by a linear weighting of the individual component functions (e.g., a mixture of 1 SCCM of CO and 3 SCCM of $O_2$ would produce a pressure equal to $P_{total} = 0.25 \times P_{CO}(1\ SCCM) + 0.75 \times P_{O_2}(1\ SCCM)$.

A parabolic equation for $CO_2$ (blue dashed line) was generated through an extrapolation from the experimental data for CO, $O_2$, and additional data for $H_2$. This process is shown in Figure S4b. The kinetic theory of gases states that the flux of an ideal gas through a small opening (*e.g.*, the aperture outlets) is inversely proportional to the square root of the molar mass of that gas. Since the gases here can be considered ideal, it is possible to normalize the flow rate by the molar mass of each species and to extrapolate the parabola for $CO_2$ from that. The extrapolated function, converted back to the real units of SCCM, is plotted as the dashed blue line in Figure S4a.

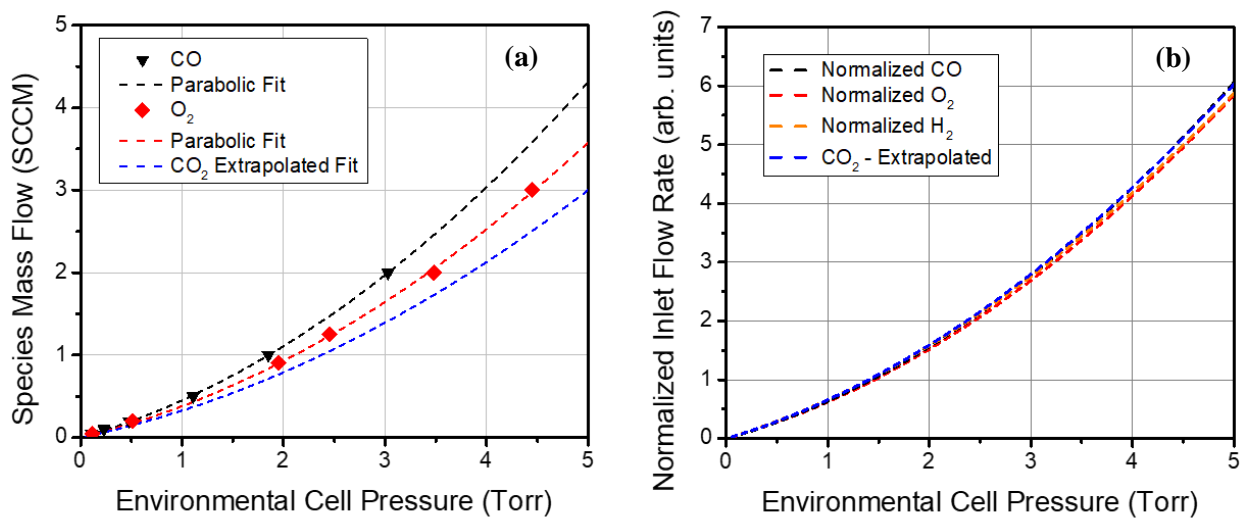

**Figure S4.** **(a)** Experimental, fitted, and extrapolated pressure-flow rate curves for CO (black triangles), $O_2$ (red diamonds), and $CO_2$ (blue dashed line). The parabolic equations plotted in the figure were used to specify the pressure at the cell outlet for a given inlet flow rate and composition. **(b)** Extrapolation of $CO_2$ parabolic pressure-flow relationship (blue dashed line) based on normalization of experimental CO, $O_2$, and $H_2$ data.

## Appendix 5: Expressions for transport properties

Mathematical expressions for properties relevant to heat and mass transport have been located from the literature and are summarized here. Polynomial expressions for the viscosity of CO, $O_2$, and $CO_2$ were determined from published data[3–5] and are given in Equations (S4) – (S6) below:

(S4) $\mu_{CO} = -2.210 \times 10^{-11} T^2 + 5.796 \times 10^{-8} T + 2.368 \times 10^{-6}$ [Pa s]

(S5) $\mu_{O_2} = -2.485 \times 10^{-11} T^2 + 6.873 \times 10^{-8} T + 2.374 \times 10^{-6}$ [Pa s]

(S6) $\mu_{CO_2} = -1.738 \times 10^{-11} T^2 + 5.889 \times 10^{-8} T - 1.082 \times 10^{-6}$ [Pa s]

Expressions for the binary diffusivities of CO, $O_2$, and $CO_2$ are found in the low-pressure limit for CO, $O_2$, and $CO_2$[6] and are given in Equations (S7) – (S9) below:

(S7) $D_{CO,CO_2} = 5.77 \times 10^{-6} T^{1.803} \times \frac{101325}{p}$ [$cm^2\ s^{-1}$]

(S8) $D_{CO,O_2} = 1.13 \times 10^{-5} T^{1.724} \times \frac{101325}{p}$ [$cm^2\ s^{-1}$]

(S9) $D_{O_2,CO_2} = 1.56 \times 10^{-5} T^{1.661} \times \frac{101325}{p}$ [$cm^2\ s^{-1}$]

Note that $D_{ik}$ with $i = k$ is defined as 1.

Mixture-averaged diffusivities, $D_i^m$, are calculated for each component $i$ by Equation (S10):

(S10) $D_i^m = \frac{1-\omega_i}{\sum_{i \neq k} \frac{x_k}{D_{ik}}}$

Here, $D_i^m$ represents the mixture-averaged diffusivity for component $i$, $\omega_i$ represents the mass fraction of component $i$ in the mixture, $x_k$ represents the mole fraction of component $k$, and $D_{ik}$ is again the binary diffusivity for components $i$ and $k$, calculated in Equations (S7) – (S9) above.

Polynomial expressions for the heat capacities of CO, O₂, and CO₂ are located in the literature[7] and given in Equations (S11) – (S12) below:

(S11) $\frac{C_{p_{CO}}}{R} = 3.02 \times 10^{-12} T^4 - 1.08 \times 10^{-8} T^3 + 1.45 \times 10^{-5} T^2 - 8.17 \times 10^{-3} T - 2.92 \times 10^2 T^{-1} + 1.48 \times 10^4 T^{-2} + 5.72$

(S12) $\frac{C_{p_{O_2}}}{R} = 1.03 \times 10^{-12} T^4 - 2.02 \times 10^{-9} T^3 - 6.83 \times 10^{-7} T^2 + 4.29 \times 10^{-3} T + 4.84 \times 10^2 T^{-1} - 3.42 \times 10^4 T^{-2} + 1.11$

(S13) $\frac{C_{p_{CO_2}}}{R} = 2.84 \times 10^{-13} T^4 - 7.68 \times 10^{-10} T^3 - 2.12 \times 10^{-7} T^2 + 2.50 \times 10^{-3} T - 6.26 \times 10^2 T^{-1} + 4.94 \times 10^4 T^{-2} + 5.30$

Polynomial expressions for the unadjusted thermal conductivities of CO, O₂, and CO₂ are found as a function of temperature in the low-pressure limit[4,5,8] and are given in Equations (S14) – (S16) below:

(S14) $k_{0,CO} = -2.178 \times 10^{-8} T^2 + 8.817 \times 10^{-5} T + 5.410 \times 10^{-4} \ [W \ m^{-1} \ K^{-1}]$

(S15) $k_{0,O_2} = -1.161 \times 10^{-8} T^2 + 7.903 \times 10^{-5} T + 3.485 \times 10^{-3} \ [W \ m^{-1} \ K^{-1}]$

(S16) $k_{0,CO_2} = 8.309 \times 10^{-5} T - 8.121 \times 10^{-3} \ [W \ m^{-1} \ K^{-1}]$

## Appendix 6: Pellet loading profile and sensitivity analysis to $X_{CO}$

Based on observations during *operando* TEM sample preparation, it was decided to model the loading of catalyst in the *operando* pellet with an egg-shell profile, whereby the majority of catalyst is located near the pellet's surface. We incorporate the loading profile into the model by introducing a spatial dependence on the magnitude of *A*, the pre-exponential factor of the reaction rate constant. All else equal, the rate of a catalytic reaction scales proportionally with the mass of catalyst involved, so factoring the loading profile into the linear term of the rate constant offers a simple way to describe the distribution of catalyst loaded into the *operando* pellet reactor. We describe the egg-shell distribution with an exponential function that depends on the depth of penetration into the pellet, as described in Equation (S17) below:

(S17) $\quad A(d) = A_0 * e^{-\gamma * d}$

Here, *A(d)* represents the pre-exponential factor with a spatial dependence, *d* represents the depth of penetration into the pellet, $A_0$ represents a constant that governs the overall magnitude of the pre-exponential factor, and *γ* is a parameter that controls the shape of the loading profile.

The penetration depth *d* at a given point within the pellet was calculated as the distance from that point to the nearest external surface of the pellet. It is possible to define a shell thickness for the egg-shell catalyst loading profile. Here, we define the shell thickness to be the distance into the *operando* pellet throughout which $1 - \frac{1}{e} \cong 63\%$ of the catalyst mass is contained. Visualizations of the catalyst loading profile for shell thicknesses ranging from 28 – 300 μm are provided in Figure S5(a)-(d) below. As the shell thickness was varied, the integrated magnitude of *A* was normalized in order to maintain the same amount of catalyst in all cases.

A sensitivity analysis was performed to evaluate the impact the uniformity of the catalyst loading profile had on the distribution of catalytically produced $CO_2$. Steady state simulations were performed for an inlet reactant gas mixture of ~2.2 Torr of stoichiometric CO and $O_2$ with a furnace thermocouple set point of 200 – 270 °C. Note that this corresponds to a condition when the catalyst is active and producing $CO_2$. The thickness of the catalyst shell was varied across an order of magnitude from 28 – 300 µm, while the mole fraction of catalytically produced $CO_2$ was investigated along the EELS line, at the TEM grid, and in the *operando* pellet. The mole fraction of catalytically-produced $CO_2$ along the EELS line, at the TEM grid, and in the *operando* pellet is plotted as a function of the catalyst shell thickness in Figure S6.

Varying the thickness of the catalyst shell from 28 – 300 µm had a weak impact on the $CO_2$ mole fraction for the two temperatures explored here. For simulations conducted at 270 °C, as the shell thickness increased from 28 to 300, the value of the $CO_2$ mole fraction along the EELS line increased from 0.571 to 0.589 (by 0.018, or 3.1%). At 200 °C, the effect is less and the $CO_2$ mole fraction along the EELS line changes by about 1.2%. As seen in the graph by comparing the top and bottom curves, variations of 1 – 3% are far smaller than the changes that result from ramping up the furnace temperature. Thus, we conclude that the interpretation of the results presented in the main text are not sensitive to slight changes in the distribution of the catalyst loaded in the reactor. It is worth noting that a shell thickness of 40 µm was used in the simulations presented in the main text.

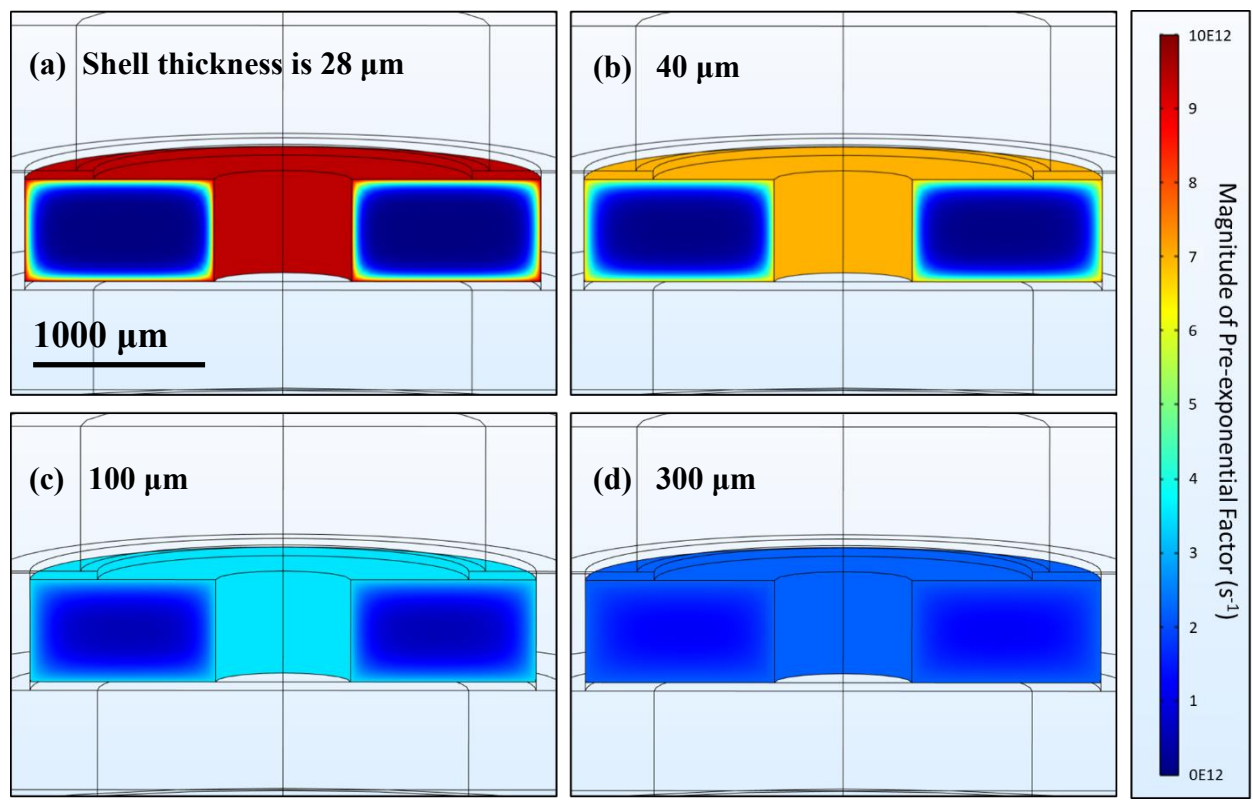

**Figure S5**. Surface plots of the three-dimensional loading profiles of catalyst in the *operando* pellet that give catalyst shell thickness of **(a)** 28 μm, **(b)** 40 μm, **(c)** 100 μm, and **(d)** 300 μm. The quantity plotted is the magnitude of the spatially varying pre-exponential factor, $A(d)$, with the legend displaying units scaled to 1E12 s$^{-1}$. A spatial scale bar of 1000 μm is provided in **(a)**. Note that as the shell thickness was changed, the integrated magnitude of $A$ throughout the pellet was normalized so the total amount of catalyst modeled in the reactor remained the same. Also note that the profile for a shell thickness of 300 μm yields a catalyst loading that is highly uniform.

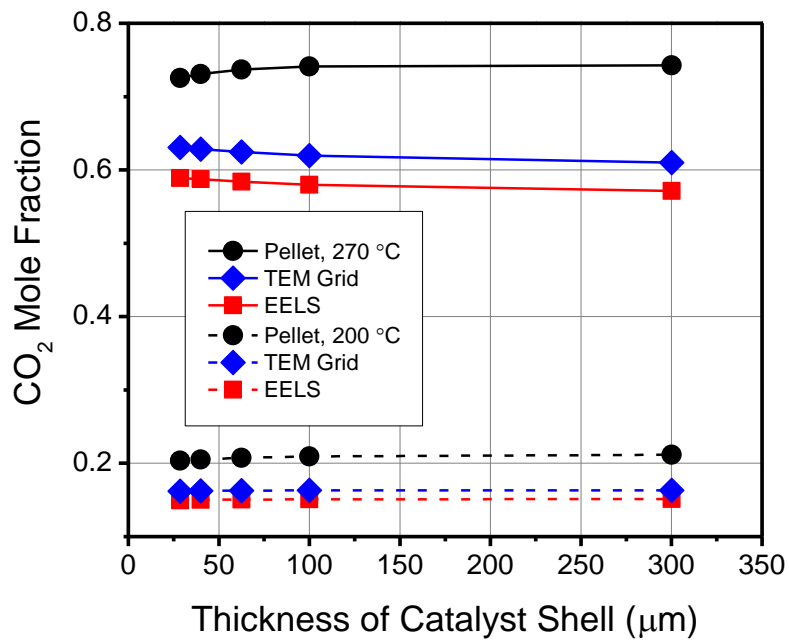

**Figure S6**. Effect of varying the catalyst shell thickness on the mole fraction of catalytically-produced $CO_2$ in the *operando* pellet (black circles), at the TEM grid (blue diamonds), and along the EELS line (red squares). Results are shown for simulations conducted at both 270 °C (solid lines) and 200 °C (dashed lines). Note that a shell thickness of 40 µm was used in the simulations presented in the main text.

## Appendix 7: Porosity sensitivity analysis

Given that the porosity was estimated by means of measuring the actual density of the pellet against the theoretical density of pure borosilicate glass, a sensitivity analysis was undertaken to probe the effect of changes in the porosity on the distribution of catalytically produced $CO_2$. An experimental measurement of the porosity was attempted with $N_2$ adsorption, but the measurement did not yield reliable results as the pores were much too large. To conduct the sensitivity analysis in the model, steady state simulations were performed for an inlet reactant gas mixture of ~2.2 Torr of stoichiometric CO and $O_2$ with a furnace thermocouple set point of 200 – 270 °C. Note that this corresponds to a condition when the catalyst is active and producing $CO_2$. The porosity of the *operando* pellet was varied from 0.5 – 0.8 while the mole fraction of catalytically produced $CO_2$ was investigated along the EELS line, at the TEM grid, and in the *operando* pellet. The results are plotted as a function of porosity in Figure S7 below.

It can be seen from the figure that varying the porosity of the *operando* pellet by physically reasonable values of 0.1 – 0.2 did not alter the $CO_2$ mole fraction by more than 1-2% for the two temperatures explored here. For simulations conducted at 270 °C, as the porosity increases from 0.5 to 0.8, the value of the $CO_2$ mole fraction along the EELS line increases from 0.592 to 0.582 (by 0.1, or 1.8%). At 200 °C, the effect is even less pronounced and the $CO_2$ mole fraction along the EELS line changes by about 0.3%. Measured at other locations, the $CO_2$ mole fraction exhibits less dependence on the porosity of the pellet, changing by fewer than 1.5%, as seen in the graph. Importantly, the overall trends of the data remain the same across the porosity parameter space. Thus, for physically reasonable variations in the estimated value of the porosity that range from 0.1 – 0.2, there were not significantly different outcomes in the behavior or interpretation of the model, and therefore the estimated value of $\varepsilon = 0.7$ can be treated as acceptable.

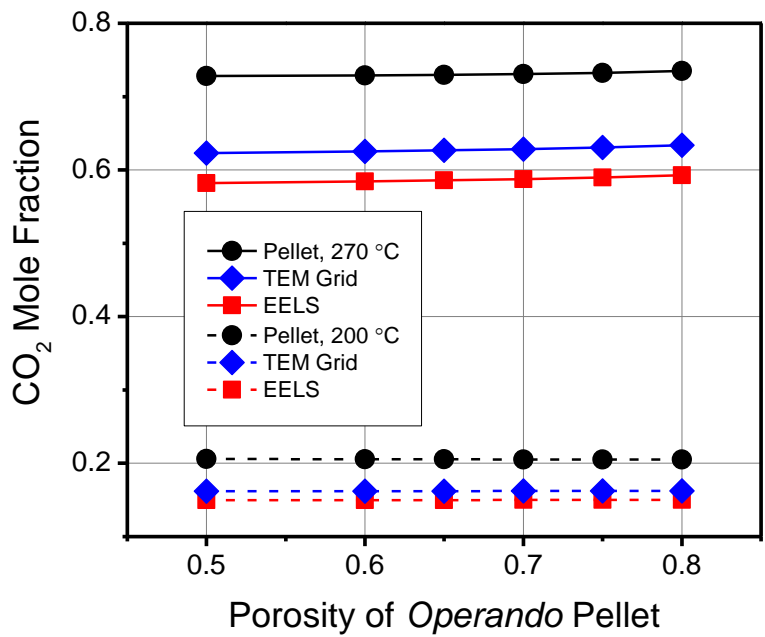

**Figure S7**. Effect of varying *operando* pellet porosity on the mole fraction of catalytically-produced $CO_2$ in the *operando* pellet (black circles), at the TEM grid (blue diamonds), and along the EELS line (red squares). Results are shown for simulations conducted at both 270 °C (solid lines) and 200 °C (dashed lines). Note that the value of the porosity used in the simulations presented in the main text is $\varepsilon = 0.7$.

## Appendix 8: Deriving CO conversion in terms of CO₂ mole fraction

Recall from Equation (15) in the main text that the CO conversion $X_{CO}$ is defined as:

(S18) $\quad X_{CO} = \dfrac{\dot{n}_{CO,in} - \dot{n}_{CO,out}}{\dot{n}_{CO,in}}$

Where $\dot{n}_{CO,in}$ is the molar flow rate (mol/s) of CO into the ETEM chamber, which is calculated in the model by taking a surface integral of the molar flux of CO across the inlet surface into the ETEM. The variable $\dot{n}_{CO,out}$ is the molar flow rate of CO out of the ETEM, which is calculated similarly by taking a surface integral of the molar flux of CO across the pumping aperture outlets.

Notice that Equation (S18) can be reformulated as a carbon balance – i.e., either C is in CO or it is in CO₂:

(S19) $\quad X_{CO} = \dfrac{\dot{n}_{CO2,out}}{\dot{n}_{CO,out} + \dot{n}_{CO2,out}}$

By dividing both halves of the fraction by the total outlet molar flow rate, $\dot{n}_{total,out}$, we obtain the CO conversion in terms of the mole fraction of CO, $y_{CO}$, and the mole fraction of CO₂, $y_{CO2}$:

(S20) $\quad X_{CO} = \dfrac{\dot{n}_{CO2,out}}{\dot{n}_{CO,out} + \dot{n}_{CO2,out}} \div \dfrac{\dot{n}_{total,out}}{\dot{n}_{total,out}} = \dfrac{y_{CO2}}{y_{CO} + y_{CO2}}$

Here, the mole fraction of species $i$ is defined in a standard way as the number of moles of that species divided by the total number of moles of all species. For example, the mole fraction of CO₂ within a volume of interest may be calculated as:

(S21) $\quad y_{CO2} = \dfrac{n_{CO2}}{n_{total}} = \dfrac{n_{CO2}}{n_{CO} + n_{O2} + n_{CO2}}$

For a stoichiometric mixture of reactants, $y_{CO}$ can be expressed in terms of $y_{CO2}$ by the equation:

(S22) $\quad y_{CO} = 1 - y_{CO2} - y_{O2} = 1 - y_{CO2} - \tfrac{1}{2} y_{CO} = (1 - y_{CO2}) \times \tfrac{2}{3}$

Substituting this into Equation (S20), we have that the CO conversion in terms of $y_{CO2}$ is:

(S23) $\quad X_{CO} = \dfrac{y_{CO2}}{(1 - y_{CO2}) \times \tfrac{2}{3} + y_{CO2}}$

# Supplemental References